\documentclass[usenatbib]{mn2e}

\usepackage{graphicx}
\usepackage{subfigure}

\begin{document}

\title{Modelling circumstellar discs with 3D radiation hydrodynamics}
\author[David M. Acreman, Tim J. Harries and David A. Rundle]{David M. Acreman, Tim J. Harries and David A. Rundle\\
School of Physics, University of Exeter, Stocker Road, Exeter EX4
4QL. E-mail acreman@astro.ex.ac.uk
}

\maketitle

\begin{abstract}

  We present results from combining a grid-based radiative transfer
  code with a Smoothed Particle Hydrodynamics code to produce a
  flexible system for modelling radiation hydrodynamics. We use a
  benchmark model of a circumstellar disc to determine a robust method
  for constructing a gridded density distribution from SPH
  particles. The benchmark disc is then used to determine the accuracy
  of the radiative transfer results. We find that the SED and 
  the temperature distribution within the disc are sensitive
  to the representation of the disc inner edge, which depends
  critically on both the grid and SPH resolution. The code is then
  used to model a circumstellar disc around a T-Tauri star. As the
  disc adjusts towards equilibrium vertical motions in the disc are
  induced resulting in scale height enhancements which intercept
  radiation from the central star. Vertical transport of radiation
  enables these perturbations to influence the mid-plane temperature
  of the disc. The vertical motions decay over time and the disc
  ultimately reaches a state of simultaneous hydrostatic and radiative
  equilibrium.

\end{abstract}


\section{Introduction}

Hydrodynamics calculations are used in investigations of  a variety of problems associated with star formation, such as the collapse of protostellar cores (e.g. \citealt{goodwin_2004}), the fragmentation of protostellar discs (e.g. \citealt{boss_1997}), and on the largest scale the gravo-turbulent collapse of a molecular cloud to form stars (e.g. \citealt{bate_2003, bate_2005}). 

Sophisticated as such calculations appear, the underlying physics regarding the thermal properties of the gas is much simplified. One of the usual approximations is the use of an equation of state (eos) in determining the pressure of the gas. An isothermal eos is often a good approximation in the most rarified regions, where the material is able to radiate, and thus cool, freely. The gas may be compressed adiabatically in the highest density regions, and a barotropic eos is sometimes adopted, with the polytropic exponent switching at a critical density in an attempt to reproduce the results of one-dimensional collapse calculations that incorporate radiative-transfer \citep{masunaga_1998}. The crucial point here is that the energy dumped into the cluster from accretion onto, and gravitation contraction of, the newly formed stars is neglected, and the radiation emitted by compressed gas (the `$P\,dV$' term) is assumed to  escape without further interaction.

The solution is to include radiation-transfer self-consistently  and
perform a radiation-hydrodynamics (RHD) calculation. The simplest way
to do this is to assume that the energy is transported by radiative
diffusion of photons. The diffusion approximation is a good one at
high optical depth, but breaks down in optically thin regions since
the diffusion speed may exceed the speed of light as the
mean-free-path tends to infinity; an essentially arbitrary flux
limiter is adopted in this regime. Furthermore the approximation is
often implemented as a grey method, typically using Rosseland mean
opacities tabulated as a function of temperature. Finally the
diffusion approximation cannot treat the effects of shadowing, as the radiation field may diffuse around optically thick obstacles. The crucial advantage of the diffusion approximation though is of course speed: the radiation transport step is straightforward to implement and is not dominant in the RHD calculation. The first cluster collapse and disc fragmentation calculations using flux-limited diffusion have just started appearing in the literature \citep{boss_2008, bate_2009b, offner_2009, stamatellos_2009}. The fundamental result of including a more complete thermal treatment is broadly in line with prior expectations i.e.  the heating of the gas inhibits the fragmentation process.

Meanwhile dedicated radiation transfer (RT) codes  have progressed far beyond the diffusion approximation and now take into account a wide array of processes such as polychromatic radiation transfer incorporating the radiation's polarization state, multiple scattering, Mie phase matrices etc. These sophisticated RT codes have been employed to model cores (e.g. \citealt{stamatellos_2005}), discs (reviewed by \citealt{dullemond_2007}) and even entire clusters \citep{kurosawa_2004}. The principle barrier to using these codes in RHD is that of  of computation time, but the increase in CPU power and the  adoption of parallelization mean that it is now feasible to conduct a full treatment of RT along with the hydrodynamics, thereby dropping the diffusion approximation.

This paper presents a method of coupling the {\sc torus}
radiative transfer code \citep{harries_2000} with the smoothed
particle hydrodynamics (SPH) code of \cite{bate_2009}.  {\sc torus}
performs radiative transfer calculations using the Monte-Carlo method
of \cite{lucy_1999} with a diffusion approximation in regions of high
optical depth. The radiative transfer calculations are performed on a
grid which uses adaptive mesh refinement (AMR) to allow variable
resolution.  The SPH method is well suited to dealing with the large
range in size scales encountered during star formation processes,
hence an SPH code coupled to a Monte-Carlo radiative transfer code is
a very flexible method for performing star formation calculations
incorporating radiative feedback.

Accurately coupling the radiative transfer component to the SPH
component requires an effective transformation from the Lagrangian
particle description of SPH to the Eulerian grid description of an AMR
grid. The procedure used for this transformation is tested using a
circumstellar disc benchmark (see
Sections~\ref{section:description_of_benchmark}
and~\ref{section:grid_generation}). The temperature distribution and spectral energy distribution
 of the benchmark disc are examined in
Section~\ref{section:seds} in order to investigate how well the
radiative transfer calculation works with the gridded density field
and to understand the effects of grid and SPH spatial resolution. Finally the hybrid code is 
tested under realistic conditions via a simulation of a circumstellar disc around a 
T Tauri star (Section~\ref{section:sph_disc}).

\section{Circumstellar disc benchmark}
\label{section:description_of_benchmark}

The accuracy of the SPH method has previously been examined in detail
(e.g. \cite{price_2008b} and references therein) so we do not attempt
to benchmark the SPH component of the code here. Likewise the  {\sc torus}
radiative transfer code has also been benchmarked elsewhere
\citep{pinte_2009}. Consequently we can be confident
in the accuracy of the SPH and radiative transfer codes individually and we instead
 focus on the accuracy of the mechanism by which they are
coupled. The benchmark case presented here is therefore a `static'
benchmark in which we represent an analytical density distribution
with SPH particles and test whether the radiative transfer code can
calculate accurate results using this SPH density representation.

In the absence of an analytical test case, benchmarking of codes used for calculating radiative transfer in
circumstellar discs is carried out by code comparison studies
\citep{pascucci_2004,pinte_2009}.
We base our SPH disc benchmark on the axisymmetric circumstellar disc of
\cite{pascucci_2004}.  Representing a structure with rotational
symmetry using a 3D geometry allows us to validate the operation of
our method against well tested results from other codes while fully
exercising the 3D capabilities of our own system. The geometry is a
stringent test of the gridding method, since it contains a large range
in linear scales that need to be resolved, and sharp density gradients
(in particular the disc inner edge).

The density of the disc benchmark described by \cite{pascucci_2004} is
given by
\begin{equation}
\rho\left(r,z\right) = \rho_0 \left( \frac{r}{r_d} \right)^{-1} \exp
\left( - \frac{\pi}{4} \left( \frac{z}{h\left(r \right)}  \right)^2 \right)
\label{eqn:pascucci_density}
\end{equation}
where $r$ is the distance from the central star in the mid-plane and
$z$ is the height from the mid-plane. The expression $h\left( r
\right)$ is given by
\begin{equation}
h\left( r \right) = z_d \left( \frac{r}{r_d}\right)^{1.125}  
\end{equation}
and $\rho_0$, $z_d$ and $r_d$ are constant parameters.  
{The definition of $h\left( r \right)$ is the definition used 
  by \cite{pascucci_2004} which differs from the standard scale
  height definition (used later in this paper) by a factor of $\frac{\pi}{2}$.
We base our
benchmark model on the most optically thick case presented by
\cite{pascucci_2004} which has a mid-plane optical depth of $\tau=100$
at $\lambda=550\,\rm{nm}$ and a mass of 0.011\,$\rm{M}_{\sun}$. This
requires $r_d=500\,\rm{au}$, $z_d=0.25 r_d$ and
$\rho_{0}=8.1614\times10^{-18}\,\rm{g\,cm^{-3}}$, with the disc truncated at
an outer radius of 1000\,au and an inner radius of 1\,au. The
    disc is irradiated by a solar-type star with a black body spectrum.

The benchmark disc is implemented using an ensemble of equal mass SPH
particles. If all the particles have the same mass then the number
density is proportional to the mass density, hence the SPH particles
sample a probability density distribution such that the probability of
finding a particle in a given volume is proportional to the mass in
that volume as a fraction of the total mass \citep{gingold_1977}. 
Hence the analytical density function can be converted into a
probability function which describes the probability of finding a
particle in a given volume.  An ensemble of particle positions can
then be calculated by randomly sampling the probability density
functions for $r$ and $z$ and assigning a uniformly random azimuthal 
angle $\phi$. The particle's density is set to the value given
by Eqn~\ref{eqn:pascucci_density} based on the particle's position.

Each particle is assigned a smoothing length $h_{\rm{smooth}}$ given by 
\begin{equation}
h_{\rm{smooth}} = 1.2 \left( \frac{m_{\rm{part}}}{\rho_{\rm{part}}}
\right) ^{1/3}
\label{eqn:smlen}
\end{equation}
where $m_{\rm{part}}$ is the particle mass and $\rho_{\rm{part}}$ is
the particle density, according to the method of
\cite{price_2007b}. The smoothing length is used when reconstructing
properties represented by the ensemble of particles, for example the
density distribution on the radiative transfer grid (see
Section~\ref{section:grid_generation}).

\section{Generation of an adaptive grid from SPH particles}

\label{section:grid_generation}

An important step in setting up grid-based radiative transfer using a
density distribution derived from SPH particles is determining an
effective transformation from the Lagrangian particle description to
the Eulerian grid description.  The representation on grid should
preserve important properties of the SPH model (e.g. total mass) but
also needs to take account of the possibility that a radiative
transfer calculation requires the highest resolution in different
regions to a hydrodynamics calculation (e.g. to resolve opacity
gradients). This section determines a grid construction method which
allows a good degree of control over the number of cells in the grid,
and the location of the highest resolution regions, while maintaining
an accurate representation of the total disc mass.

\subsection{Grid generation method}

{\sc torus} uses an adaptive grid based on the octree method. In a 3D
geometry the initial grid comprises 8 cells (one octal) with 2 cells
in each dimension. An algorithm is applied to determine whether to
split each cell into a further 8 cells, similar to the method of
\cite{kurosawa_2001}. The properties of the adaptive grid are
determined by a combination of the algorithm used to decide when to
split a cell and the method used to assign density values to cells.

The density in a given grid cell is calculated using an exponential
kernel smooth. If the sum of the kernel weights is greater than 0.3
the density is normalised by the sum of the kernel weights, to ensure a
smooth density distribution within the interior of the disc. If the
sum of the weights is less than 0.3 then the density is not normalised
by the sum of the weights, in order to avoid numerical effects at the
free surface, as described by \cite{price_2007}. For more details of
the algorithm used to generate the AMR grid from SPH particles see
\cite{rundle_2009}. 

\subsection{Cartesian grid generation tests}

Two grid splitting conditions were tested. The first method splits a
grid cell if the mass within the cell exceeds a given limit. For this
condition the resolution is highest where the density is highest and
the grid cell size is analogous to the SPH particle smoothing
length. The second condition decides whether to split the cell based
on the fractional density difference between the most dense and least
dense SPH particles contained within the cell. The quantity
\begin{equation}
f_{\rm{split}}=\frac{\rho_{\rm{max}} - \rho_{\rm{min}}}{\rho_{\rm{max}} + \rho_{\rm{min}}}
\end{equation}
is calculated, where $\rho_{\rm{max}}$ is the highest SPH particle
density and $\rho_{\rm{min}}$ is the lowest SPH particle density, and
the cell is split if a specified value of $f_{\rm{split}}$ is
exceeded. The first condition gives a resolution like that of the
original SPH representation whereas the second condition should better
represent gradients in density which are important for radiative
transfer calculations.

To illustrate the effects of the different splitting methods two
example grids are shown in Fig.~\ref{fig:example_meshes}. A grid
constructed from $10^7$ particles, using a mass per cell limit of
$5\times10^{26}\,\rm{g}$, is plotted in
Fig.~\ref{fig:mass_condition_mesh}. This plot is a slice perpendicular
to the disc mid-plane and shows increased resolution towards the disc
centre and mid-plane where the density is higher.  A grid with
$f_{\rm{split}}=0.1$, also using $10^7$ particles, is plotted in
Fig.~\ref{fig:density_condition_mesh}. 
The resolution increases towards the edge of the disc, where there 
are large density gradients,and also increases towards the centre 
of the disc. 
When the density split
    condition is used there is a noticeable asymmetry to the grid
    resolution, due to statistical fluctuations in the SPH particle
    distribution. The effect is seen more readily with the
    density split condition because the grid resolution is being
    enhanced in lower density regions, where SPH particles sample the
    density distribution less well than in high density regions.
\begin{figure*}
  \subfigure[AMR grid with a mass per cell condition of
    $5\times10^{26}\,\rm{g}$.]
    {\includegraphics[scale=0.22]{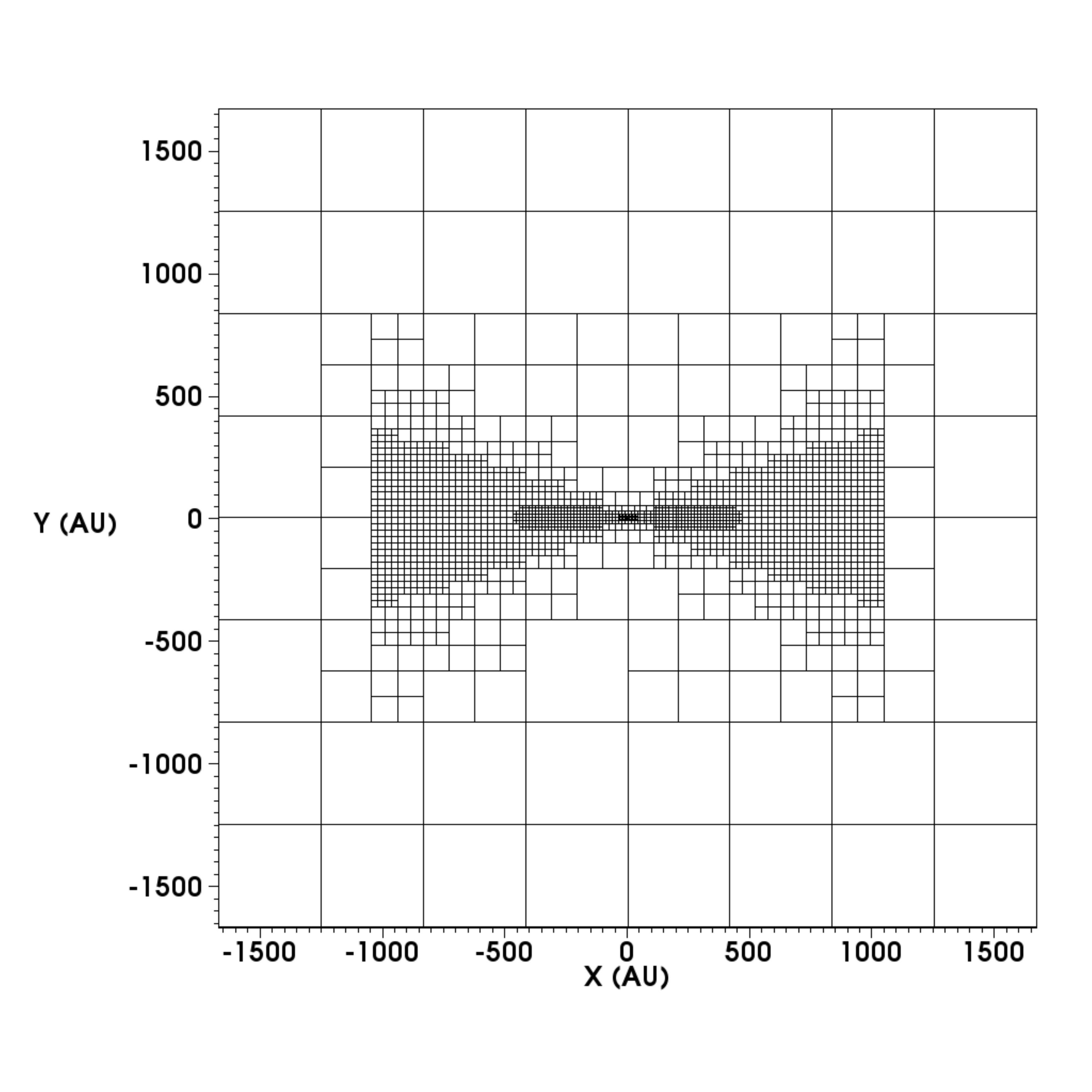}
  \label{fig:mass_condition_mesh}}
  \subfigure[AMR grid with a density condition of
    $f_{\rm{split}}=0.1$.]
    {\includegraphics[scale=0.22]{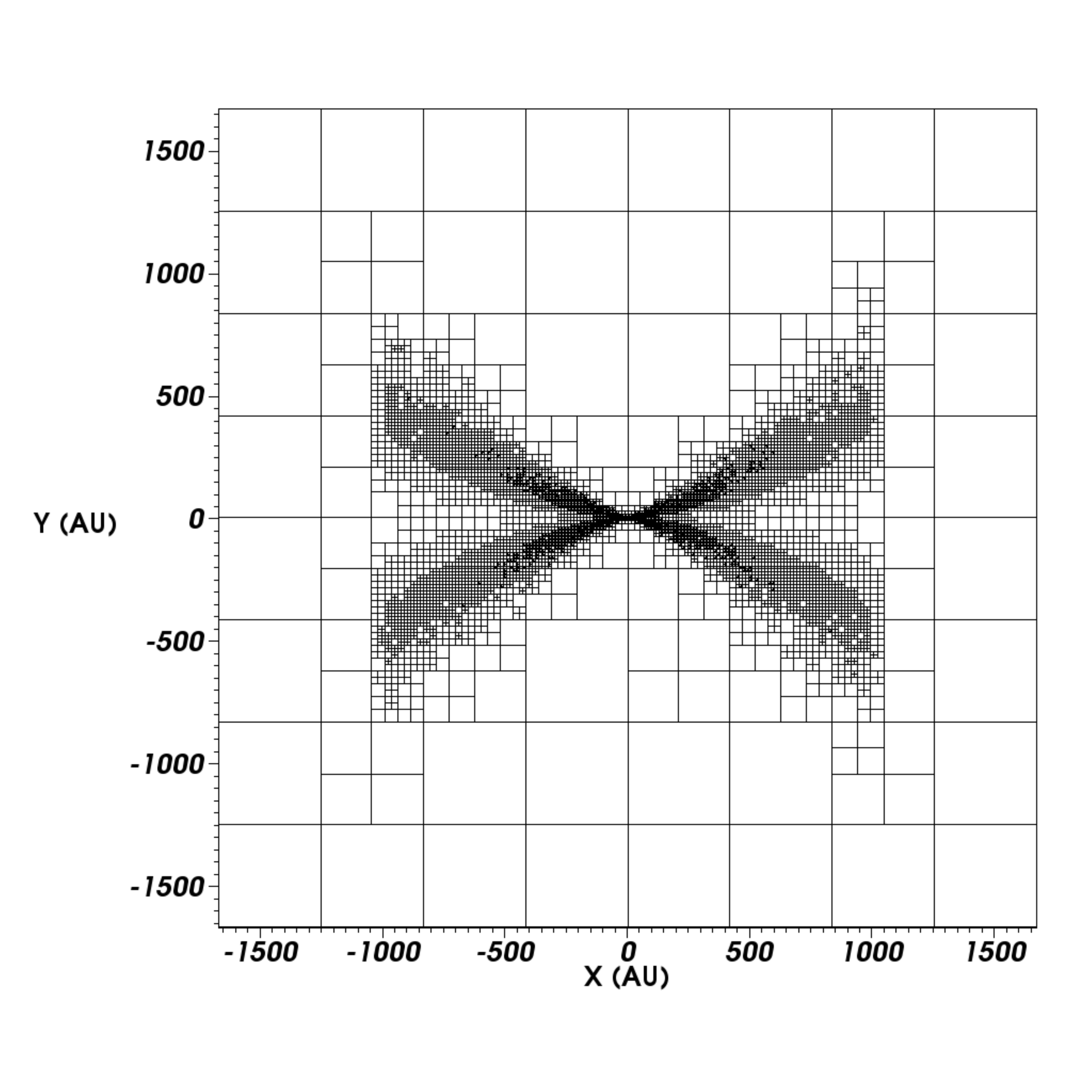}
      \label{fig:density_condition_mesh}}
  \caption{Example AMR grids for a mass per cell splitting condition
    of  $5\times10^{26}\,\rm{g}$ (left)
    and a density contrast splitting condition of $f_{\rm{split}}=0.1$
    (right) constructed
    from $10^7$ SPH particles. The grids are
    slices perpendicular to the disc mid-plane.}
  \label{fig:example_meshes}
\end{figure*}

For each splitting condition a number of AMR grids were generated
using different values of the mass per cell limit or
$f_{\rm{split}}$. This was repeated for discs represented by $10^5$,
$10^6$ and $10^7$ SPH particles. Figure~\ref{fig:grid_tests} plots the
number of octals generated as a function of mass per cell or
$f_{\rm{split}}$ (solid line), and the percentage error in the disc
mass (dashed line). The percentage error in disc mass is calculated by
comparing the total mass on the AMR grid with the known mass of
$0.011\,\rm{M}_{\sun}$ based on the analytical form of the density
distribution, and a positive value indicates that mass on the grid is
too large.
\begin{figure*}
  \subfigure[Mass per cell condition]{\includegraphics[scale=0.3]{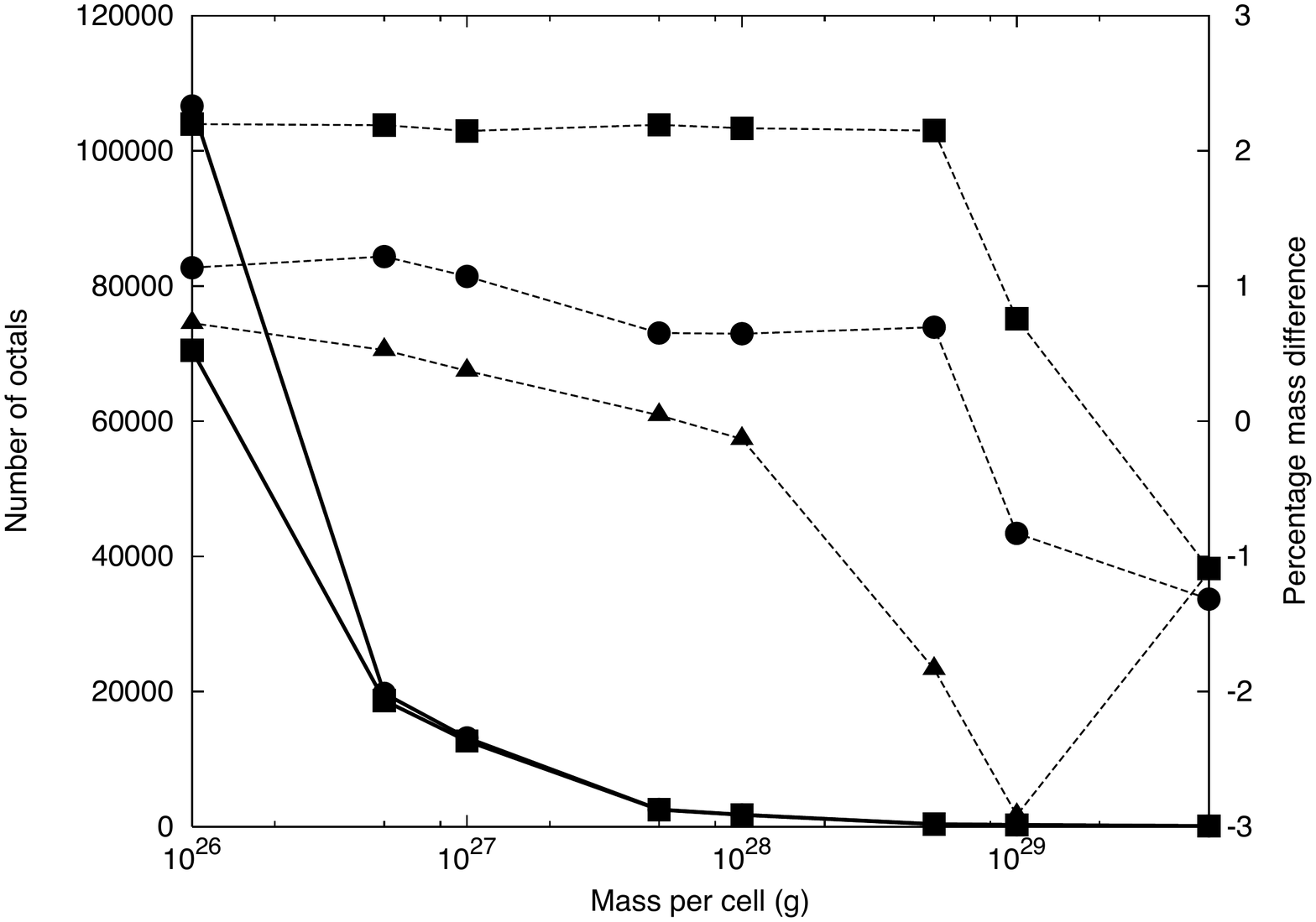}}
  \subfigure[Density contrast condition]{\includegraphics[scale=0.3]{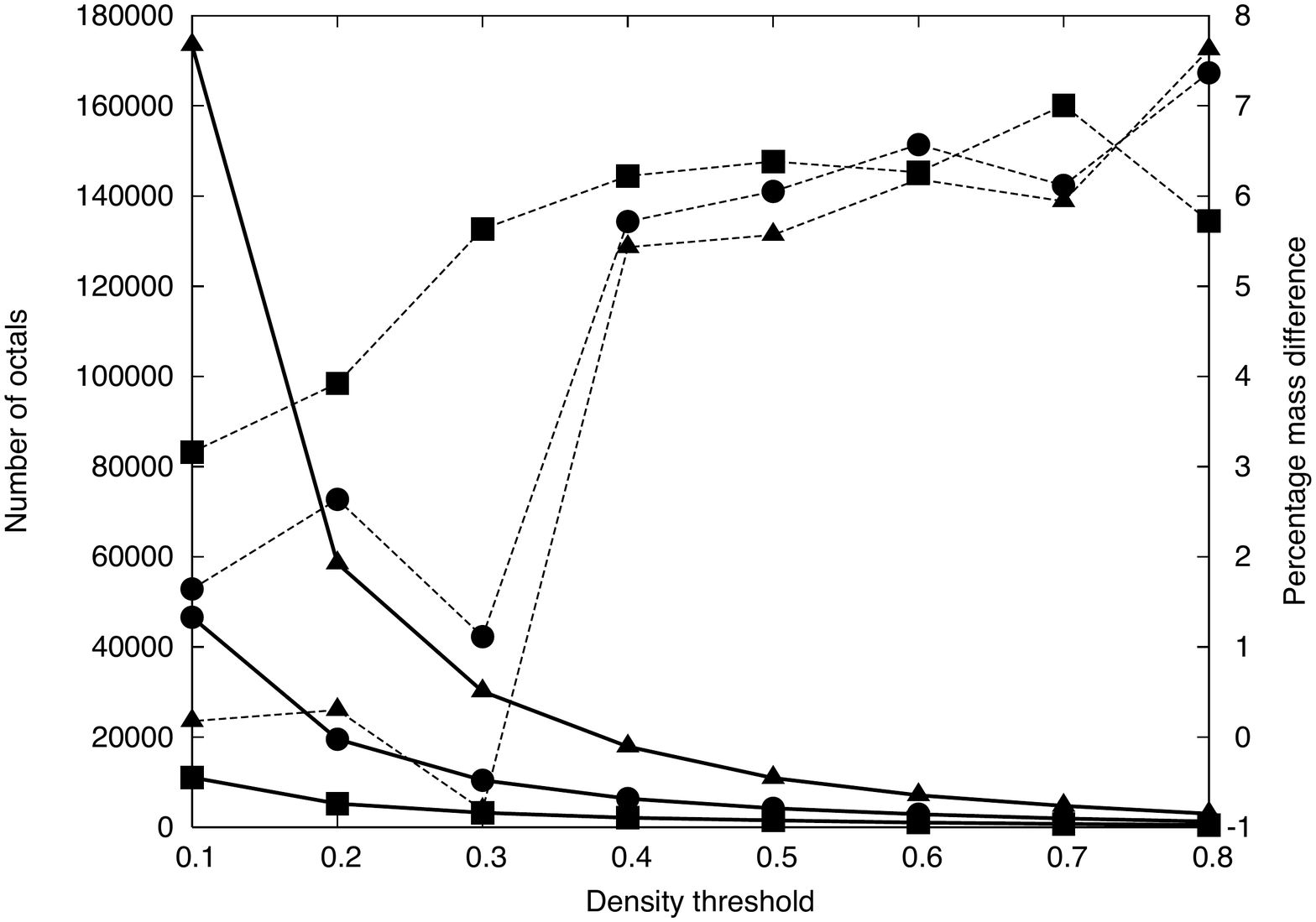}}
  \caption{Number of octals (solid line) and percentage error in total
    disc mass (dashed line) for the benchmark disc represented on
    Cartesian AMR grids. The grids were generated using either a 
    mass per cell limit (left) or density contrast limit (right) and
    different numbers of particles ($10^5$, $10^6$ or $10^7$). Results
  using $10^5$ particles are plotted with squares, $10^6$ particles
  with circles and $10^7$ particles with triangles. The number of octals for a mass
per cell condition with $10^7$ particles is not plotted as it is
indistinguishable from the case with $10^6$ particles.}
  \label{fig:grid_tests}
\end{figure*}

The mass per cell condition allows a wide range in the total number of
octals while maintaining a total mass which is correct to within a few
percent. The number of octals as a function of mass per cell limit is
consistent between runs with different numbers of SPH particles, apart
from the case with $10^5$ particles and a mass limit of
$10^{26}$\,g. In this case the mass per cell limit is smaller than the
SPH particle mass and the grid is over sampling the SPH resolution. In
the other cases the mass per cell limit is greater than the SPH
particle mass. For mass per cell limits of $10^{28}$\,g and less the
total mass error varies only slowly as a function of mass per cell and
is accurate to within a few percent. There is a positive bias in the
total mass but using more SPH particles results in a more accurate
total mass.

If a density contrast condition is used then the total number of
octals depends on the number of particles, unlike in the mass per cell
case. The general form of the number of octals as a function of
density contrast is the same in all three cases but the normalisation
varies substantially. In order to ensure a total mass accurate to with
1~per~cent the density contrast condition would require at least $10^7$
particles and a density contrast threshold of no more than 0.3.

The density contrast condition is effective in adding extra resolution
in regions of high density contrast which are likely to be important
in radiative transfer calculations. However this condition has a less
robust total mass representation than the mass per cell limit. By
using a combination of these two conditions it is possible to add
extra resolution in high contrast regions and still maintain an
accurate total mass.
 
\subsection{Cylindrical polar grid generation}

For some geometries it may be appropriate to use a cylindrical polar
co-ordinate system. A mass per cell splitting condition can still be
used but needs to be modified to take into account that for fixed
values of $dr$ and $dz$ the cell volume increases as cylindrical polar
radius increases.
Consider a Cartesian grid with cubic volume elements and let
\begin{equation}
dl_{\rm{cart}} = dx = dy = dz 
\end{equation}
so that
\begin{equation}
dl_{\rm{cart}} = \left( \frac{dM_{\rm{cart}}}{\rho_{\rm{cell}}} \right)
^{1/3}
\label{eqn:cart_cell}
\end{equation}
where $dM_{\rm{cart}}$ is the mass of a cell in the Cartesian grid and
$\rho_{\rm{cell}}$ is the density of the cell. 
In a cylindrical polar grid the mass of a cell is given by
\begin{equation}
dM_{\rm{cyl}} = \rho_{\rm{cell}} \frac{2 \pi r dr dz}{n_{\rm{az}}}
\end{equation}
if there are $n_{\rm{az}}$ uniform azimuthal cells. If the 
resolution in $r$ and $z$ is the same then we can define 
\begin{equation}
dl_{\rm{cyl}} = dr = dz 
\end{equation}
which is related to the cell mass and density by
\begin{equation}
dl_{\rm{cyl}} = \left( \frac{n_{\rm{az}} dM_{\rm{cyl}}}{2 \pi r
  \rho_{\rm{cell}}} \right)  ^{1/2}
\label{eqn:cyl_cell}
\end{equation}
If we require the resolution in cylindrical polar $r$ and $z$
to match the resolution in Cartesian $x$, $y$ and $z$ then $dl_{\rm{cyl}} =
dl_{\rm{cart}}$. From eqn~\ref{eqn:cart_cell} and \ref{eqn:cyl_cell} 
\begin{equation}
dM_{\rm{cyl}} = \frac{2 \pi r}{n_{\rm{az}}} \rho^{1/3} dM_{\rm{cart}}^{2/3}
\end{equation}
which describes the required conversion from a mass per cell limit
suitable for a Cartesian grid ($dM_{\rm{cart}}$) to an equivalent mass
per cell limit suitable for a cylindrical polar grid ($dM_{\rm{cyl}}$). 

The grid generation tests were run using a cylindrical polar geometry
with the modified mass per cell limit described above and an
unmodified density contrast condition. The results are plotted in
Fig.~\ref{fig:cp_grid_tests}.
\begin{figure*}
  \subfigure[Mass per cell condition]{\includegraphics[scale=0.3]{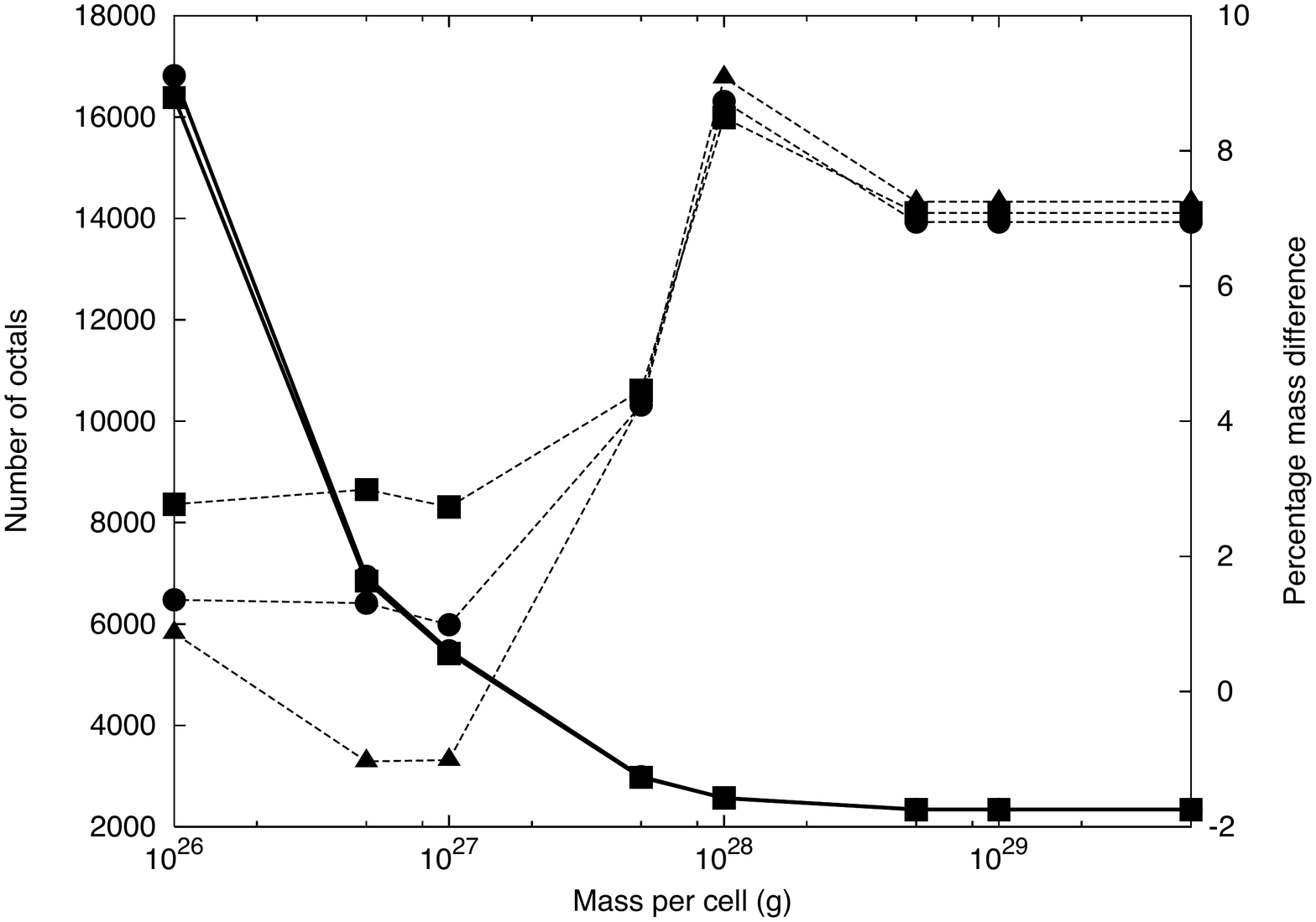}}
  \subfigure[Density contrast condition]{\includegraphics[scale=0.3]{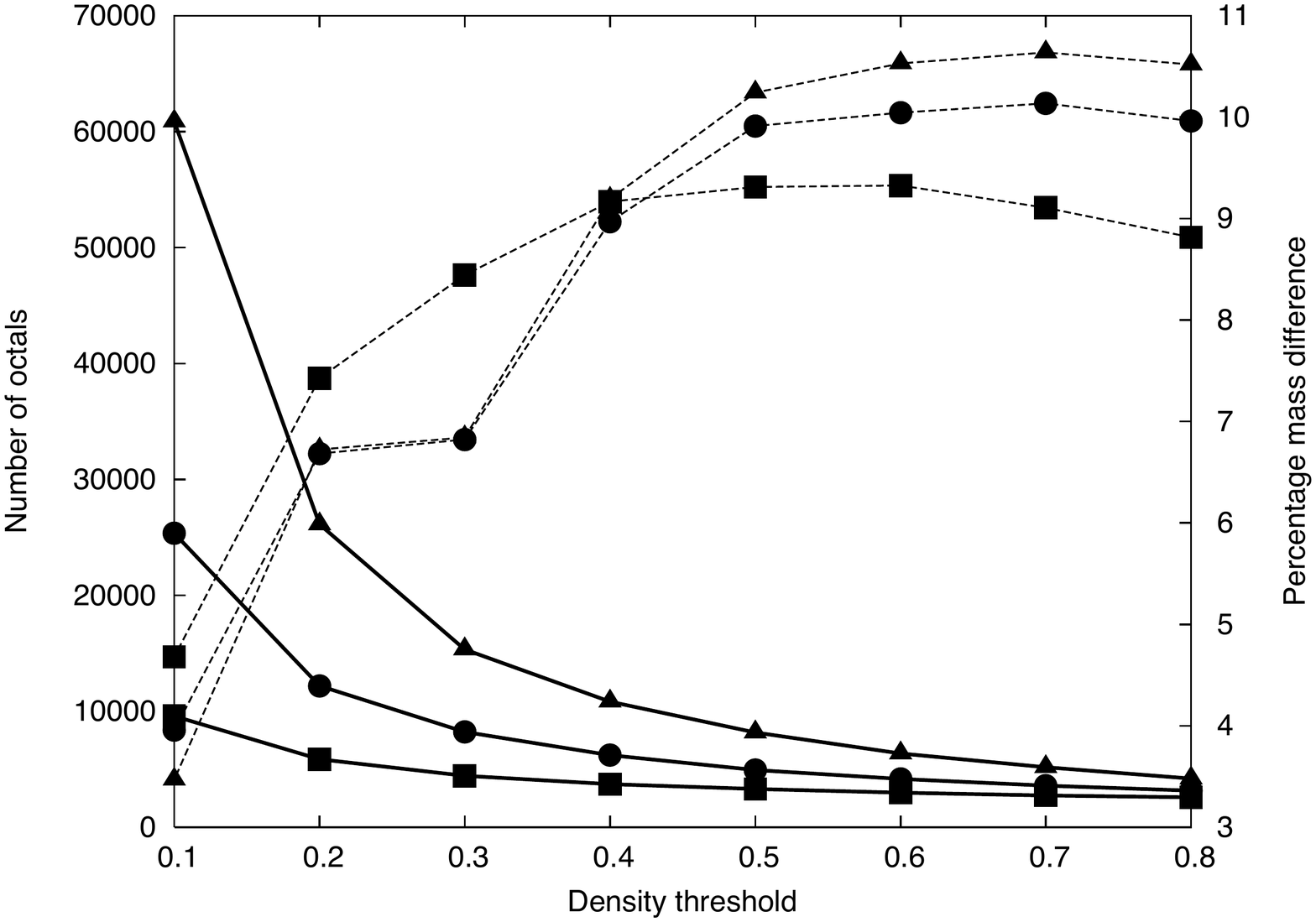}}
  \caption{Number of octals (solid line) and percentage error in total
    disc mass (dashed line) for the benchmark disc represented on
    cylindrical polar AMR grids. The grids were generated using either a 
    mass per cell limit (left) or density contrast limit (right) and
    different numbers of particles ($10^5$, $10^6$ or $10^7$). Results
  using $10^5$ particles are plotted with squares, $10^6$ particles
  with circles and $10^7$ particles with triangles. The number of octals for a mass
per cell condition with $10^7$ particles is not plotted as it is
indistinguishable from the case with $10^6$ particles.}
  \label{fig:cp_grid_tests}
\end{figure*}
Using a cylindrical polar geometry results in a significantly reduced
number of octals in the AMR grid but with larger mass errors. Provided
the mass per cell limit is no more than $10^{27}\rm{g}$ then the total
mass is well represented with significantly fewer cells than in a
Cartesian grid. The density contrast condition alone does not
represent the total mass to within 3\% accuracy even with
$f_{\rm{split}}=0.1$. Although this condition successfully adds
enhanced resolution in regions of higher density contrast it needs to
be combined with a mass per cell limit to ensure a robust total mass
representation.

\section{Comparison of temperature distribution and SEDs with
  benchmark results}

\label{section:seds}

Having investigated the procedure for setting up the radiative transfer
grid we now proceed to studying the accuracy of the results from the
radiative transfer calculation. This was carried out by comparing
temperature distributions and spectral energy distributions (SEDs)
from the benchmark disc. Temperature distribution errors provide a
clear picture of where errors in the radiative transfer calculation
are located whereas SEDs provide a direct link to an observable
quantity and are a stringent test of the accuracy of the radiative
transfer calculation. The key problem here is that the resolution necessary for adequately
describing the fluid flow is not necessarily the same as required for accurately 
modelling the radiation field.

The disc was initially modelled using $10^5$ SPH particles with an AMR
grid constructed from a mass per cell limit of $5\times10^{26}$\,g.
This number of SPH particles would normally be considered as
sufficient to adequately represent a disc of this mass in a purely
hydrodynamical calculation. The mass per cell limit was chosen in
preference to the density contrast limit to give the highest
resolution near the central region of the disc, which we expect to be
most important for determining the temperature of the disc at the
inner edge and the SED.

In Fig.~\ref{fig:nomod_tem_diff} we plot the fractional error in the
temperature distribution at radiative equilibrium for the benchmark
disc.  {\sc torus} calculates the temperature distribution using a
high resolution 2D geometry, in order to determine the correct
temperature as a function of cylindrical polar $r$ and $z$
co-ordinates. This is then compared to the temperature from the SPH
disc for each SPH particle. Positive values of the fractional error
indicate that the SPH disc is hotter than the expected value. The
temperatures show deviations of $-30$\% to $+50$\%  from the
benchmark values, a deviation that is purely driven by the resolution
of the density field derived from the SPH particles, since a `{\sc
  torus} only' calculation, where the AMR grid is refined on optical
depth criteria and the cell densities are found from the analytical
description of the density structure (Eqn.~\ref{eqn:pascucci_density}), matches the
benchmark temperatures to within 2\%. The systematic differences in
the temperature distribution are due to an increased vertical
transport of radiation from the surface layers of the disc to the
mid-plane due to a lack of spatial resolution, leading to errors in
the calculation of the specific intensities (particularly in the inner
disc region).  These systematic differences are more readily apparent
in the emergent SEDs, which are a more sensitive probe of the accuracy
of the RT calculation than the temperature distribution.

\begin{figure}
  \includegraphics[scale=0.3]{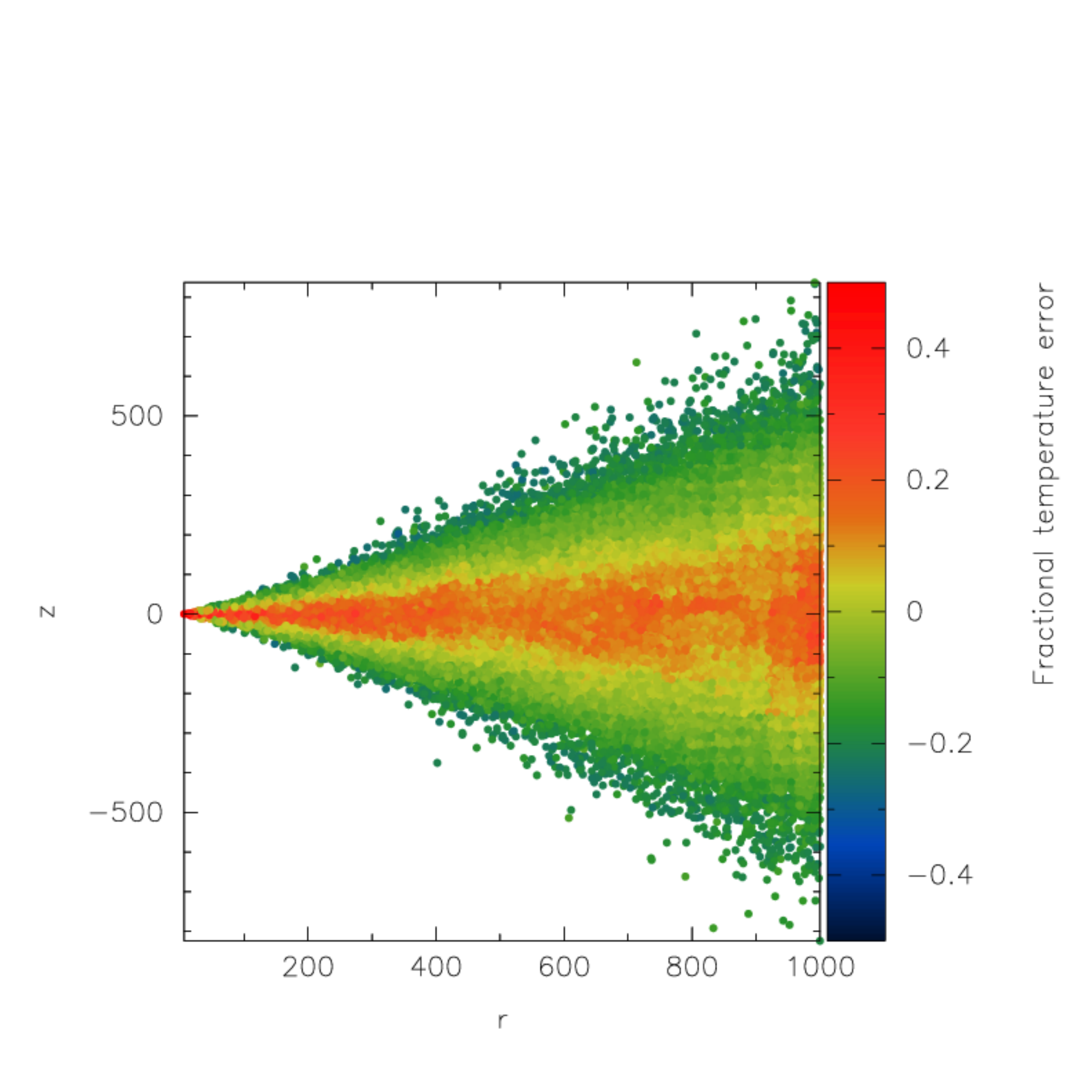}
  \caption{Fractional errors in the temperature distribution of a
    $10^5$ particle disc. The radiative transfer grid is a Cartesian
    grid generated using a mass per cell limit of
    $5\times10^{26}$\,g. Axis units are au.}
  \label{fig:nomod_tem_diff}
\end{figure}

SEDs were calculated using the {\sc torus} code for viewing
angles of 12.5 and 77.5 degrees (where an inclination angle of zero
degrees indicates that the disc is viewed face-on) and the results are
plotted in Fig.~\ref{fig:seds} (dashed line). Also plotted in
Fig.~\ref{fig:seds} are the benchmark results of \cite{pascucci_2004}
(solid line).
\begin{figure*}
  \includegraphics[scale=0.3]{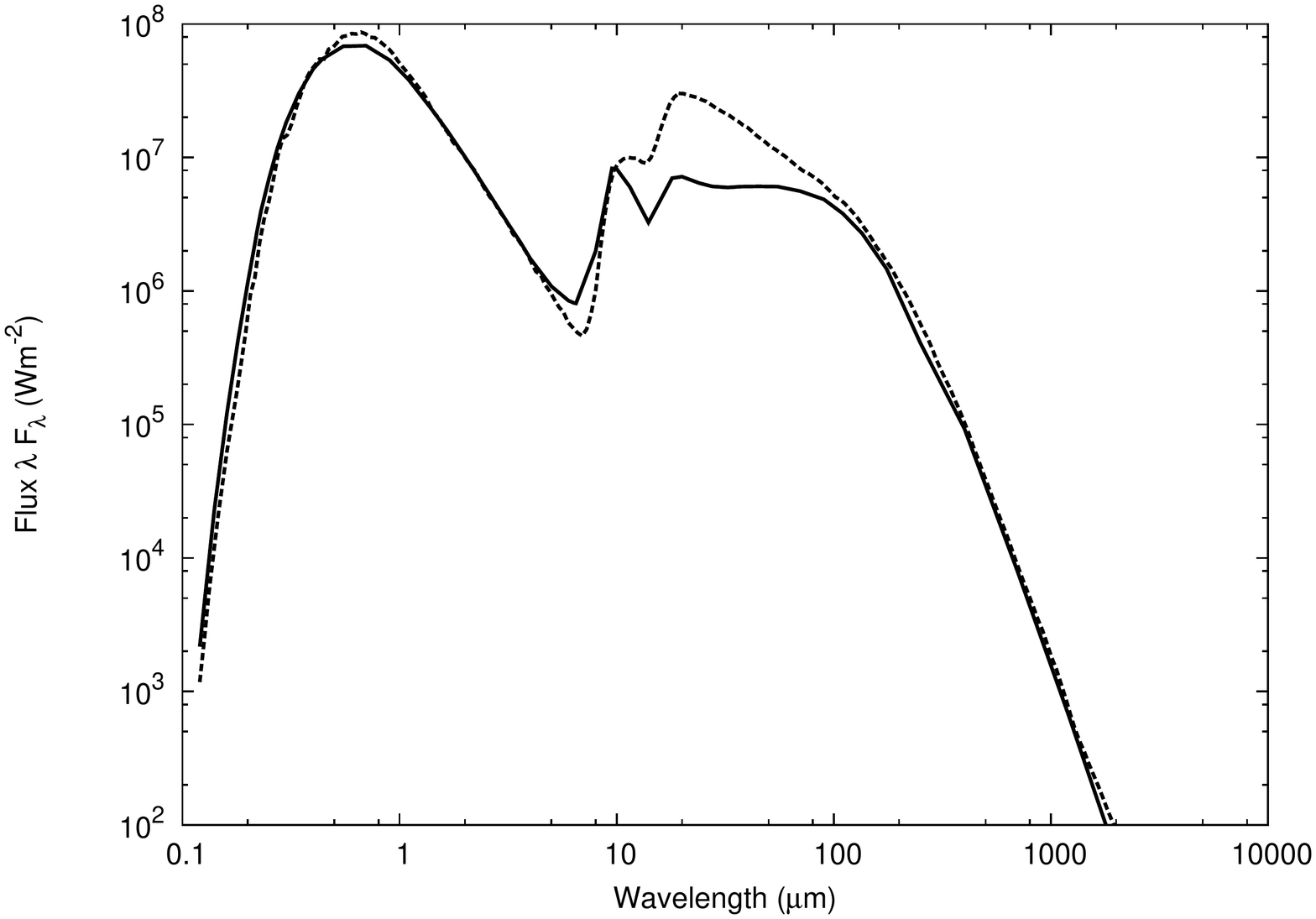}
  \includegraphics[scale=0.3]{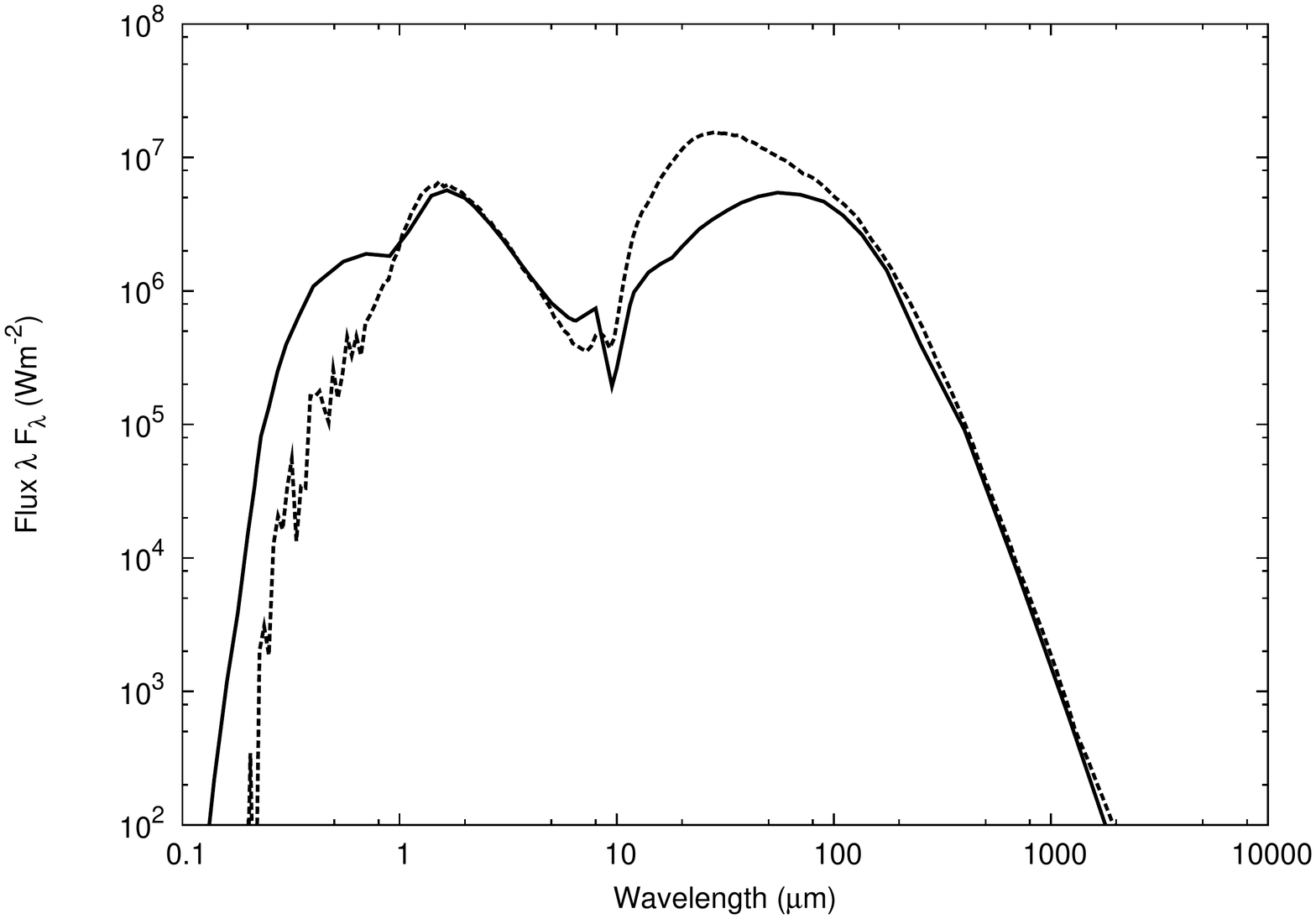}
  \caption{SEDs from the benchmark disc represented by $10^5 $ particles and a
    Cartesian radiative transfer grid generated using a mass per cell
    limit of $5\times10^{26}$\,g. SEDs are shown for inclination
    angles of 12.5 degrees (left) and 77.5 degrees (right). The
    benchmark SED is plotted as a solid line and the {\sc torus} SED
    is plotted as a dashed line. The corresponding fractional
    temperature errors are plotted in Fig.\ref{fig:nomod_tem_diff}.}
  \label{fig:seds}
\end{figure*}
Both SEDs show significant departures from the benchmark results;
there is excess emission between 10-100 $\mu\rm{m}$ and at a viewing
angle of 77.5 degrees there is also a significant reduction in flux
below 1\,$\mu\rm{m}$. Both these discrepancies are attributable to a lack of resolution within a few au of the central object. The density distribution in the central region of
this disc is plotted in Fig.~\ref{fig:basic_dens} which shows that the
central 1\,au gap is not represented. There are two effects in
operation; firstly the grid resolution in this region is larger than
1\,au  and secondly the smoothing lengths of the particles are larger
than 1\,au. In order to correctly represent the central gap there needs
to be higher grid resolution around the central source and more SPH
particles are required so that the smoothing lengths are smaller.
\begin{figure}
\includegraphics[scale=0.2]{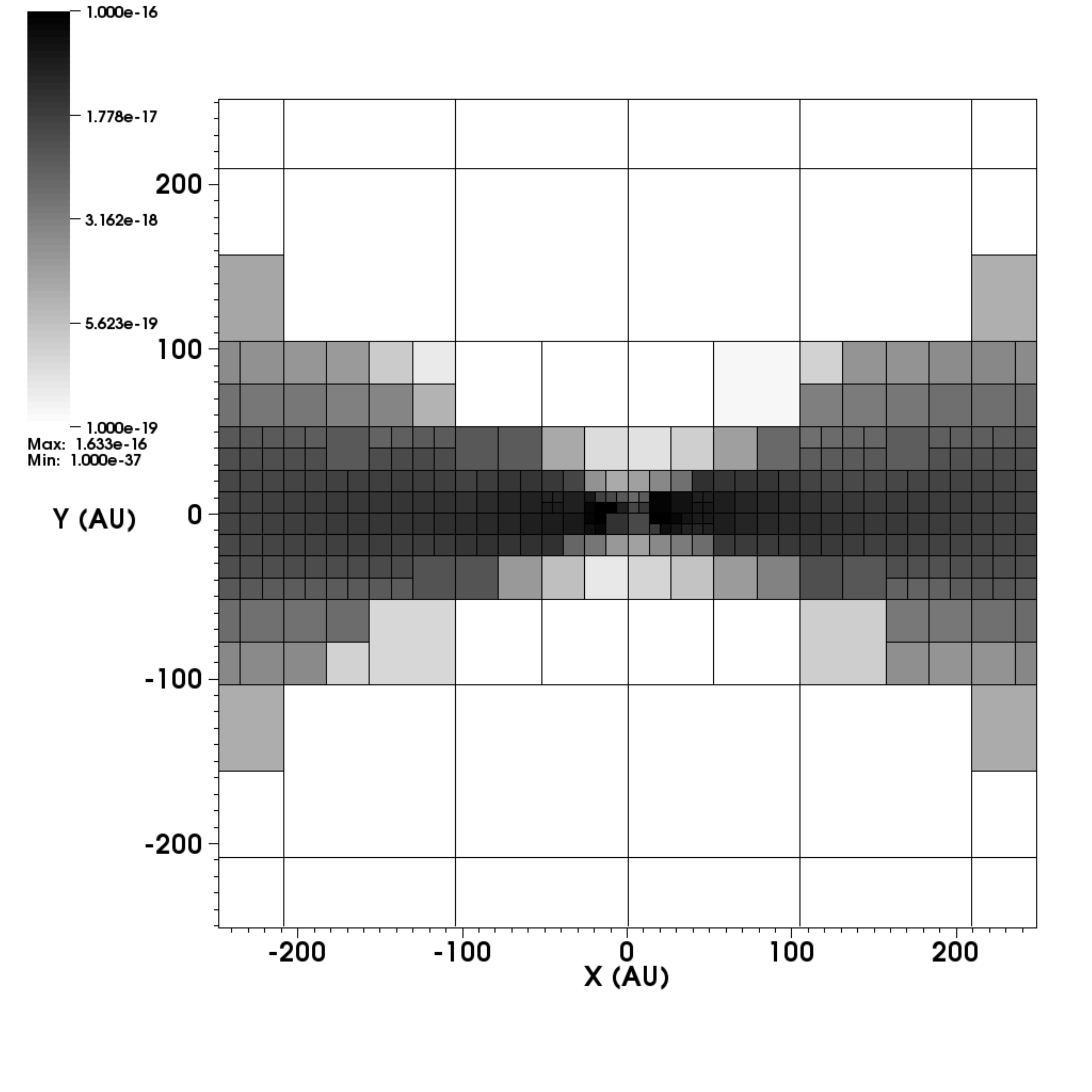}
\caption{Density ($\rm{g\,cm}^{-3}$) in the central region of the benchmark disc
  represented by $10^5$ particles and a Cartesian radiative transfer
  grid generated using a mass per cell limit of
  $5\times10^{26}$\,g.  The central 1\,au gap 
   is not represented and the effects on
  the temperature distribution and emergent SEDs are shown in
  Fig.~\ref{fig:nomod_tem_diff} and Fig.~\ref{fig:seds}
  respectively.}
\label{fig:basic_dens}
\end{figure}

\subsection{Grid resolution}
\label{subsec:grid_res}

Decreasing the mass per cell limit to achieve the required resolution
around the central source would result in too many cells being
generated in other regions of the disc. In order to achieve a grid
resolution of less than 1\,au around the source it was necessary to
specify additional criteria for refining the grid. Three cubes of
increased resolution were added to the AMR grid with sizes
$1.5\times10^{15}$\,cm, $3\times10^{14}$\,cm and
$6\times10^{13}$\,cm. Within each cube it was required that the cell
size be no larger than 0.1 of the cube dimension. In order to confirm
that the grid modifications provide sufficient resolution to generate
an accurate SED three calculations were performed with the central
part of the disc forced to the correct density values from the
analytical density description.  A spherical region with radius
$1.0\times10^{15}$\,cm, $5.0\times10^{14}$\,cm or $2.5\times10^{14}$\,cm
was over written with the benchmark disc density after the initial
grid generation was performed. The resulting SEDs are plotted in
Fig.~\ref{fig:sed_forced}. The solid line shows the benchmark result
and the other lines show SEDs for forcing regions of radius
$1.0\times10^{15}$\,cm (long dashed line), $5.0\times10^{14}$\,cm (short
dashed line) and $2.5\times10^{14}$\,cm (dotted line).
\begin{figure*}
  \includegraphics[scale=0.3]{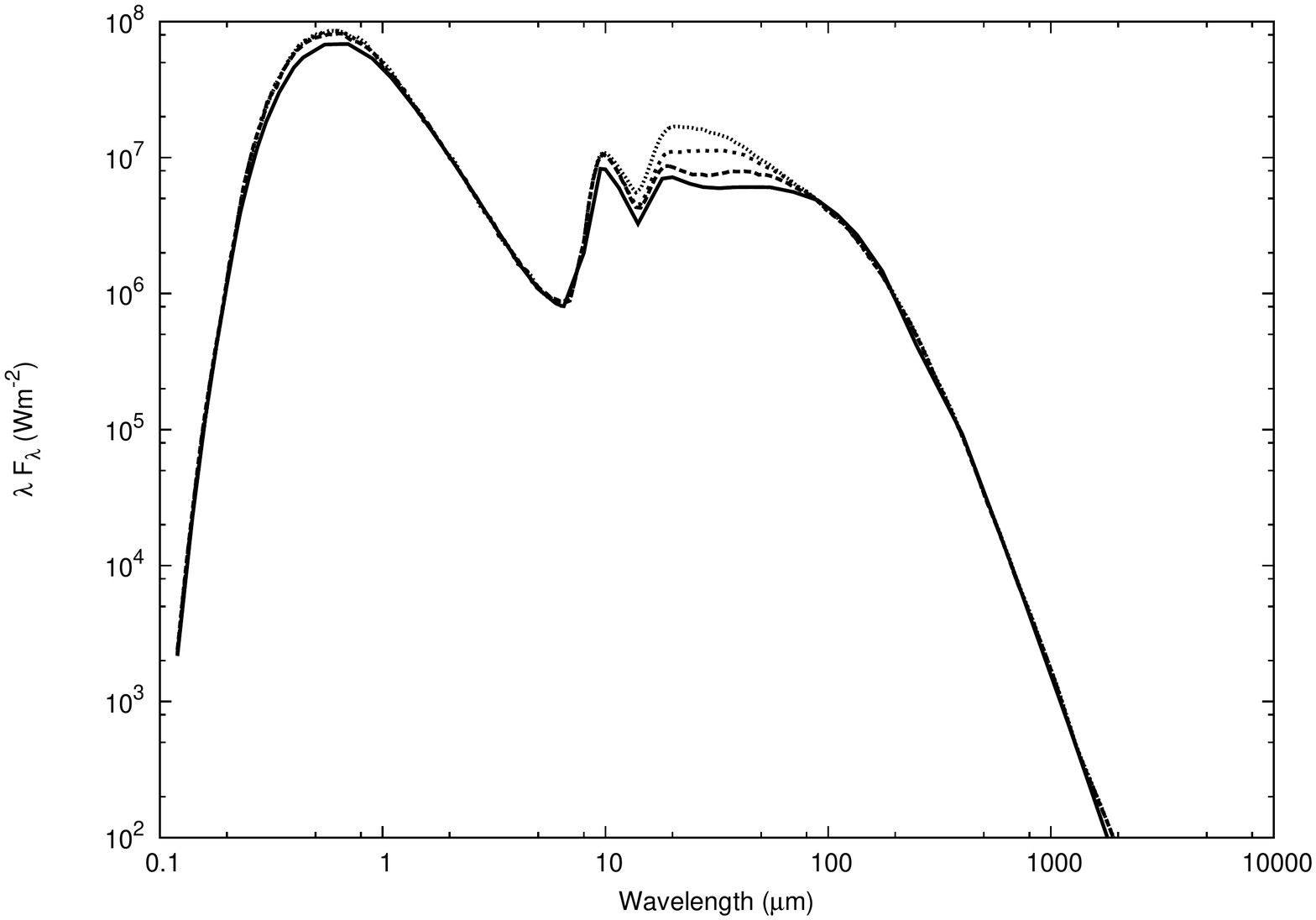}
  \includegraphics[scale=0.3]{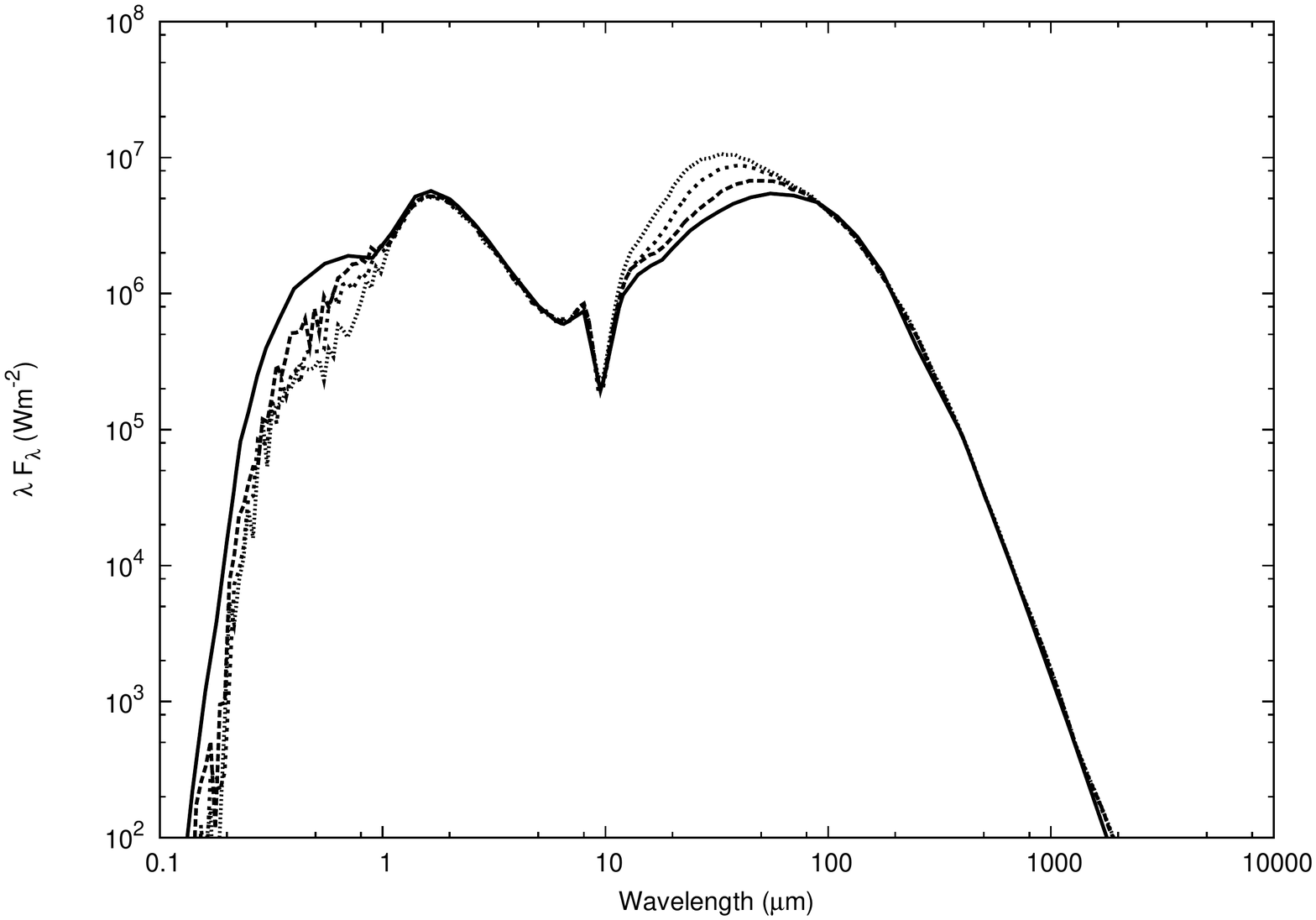}
  \caption{ SEDs from the benchmark disc represented by $10^5 $ particles and
    a Cartesian radiative transfer grid generated using a mass per
    cell limit of $5\times10^{26}$\,g. Extra resolution has been added
    around the central source and the {\sc torus} density distribution
    is forced to the correct value within a radius of
    $1.0\times10^{15}$\,cm (long dashed line), $5.0\times10^{14}$\,cm
    (short dashed line) and $2.5\times10^{14}$\,cm (dotted line). SEDs
    are shown for inclination angles of 12.5 degrees (left) and 77.5
    degrees (right) with the benchmark result plotted as a solid
    line.}
  \label{fig:sed_forced}
\end{figure*}
With the correct density forced out to a radius of
$1.0\times10^{15}$\,cm the SED is close to the benchmark result for
both viewing angles, however if the forced region is reduced in size
there is a significant discrepancy relative to the benchmark SED. This
confirms that the grid now has enough resolution to generate an
accurate SED but also indicates that using $10^5$ particles does not
give a sufficiently accurate density distribution within the central
$1.0\times10^{15}$\,cm of the disc. 

\subsection{Number of particles}
Further calculations were performed using different numbers of SPH
particles and the enhanced grid resolution described above in order to
determine the impact of the number of SPH particles on the accuracy of
the radiative transfer calculation. The density distribution in the
central region of the disc is plotted in
Fig.~\ref{fig:n_part_central_dens} in order to show how the
representation improves as the number of particles is increased.
\begin{figure*}
  \subfigure[$10^5$ particles]{\includegraphics[scale=0.22]{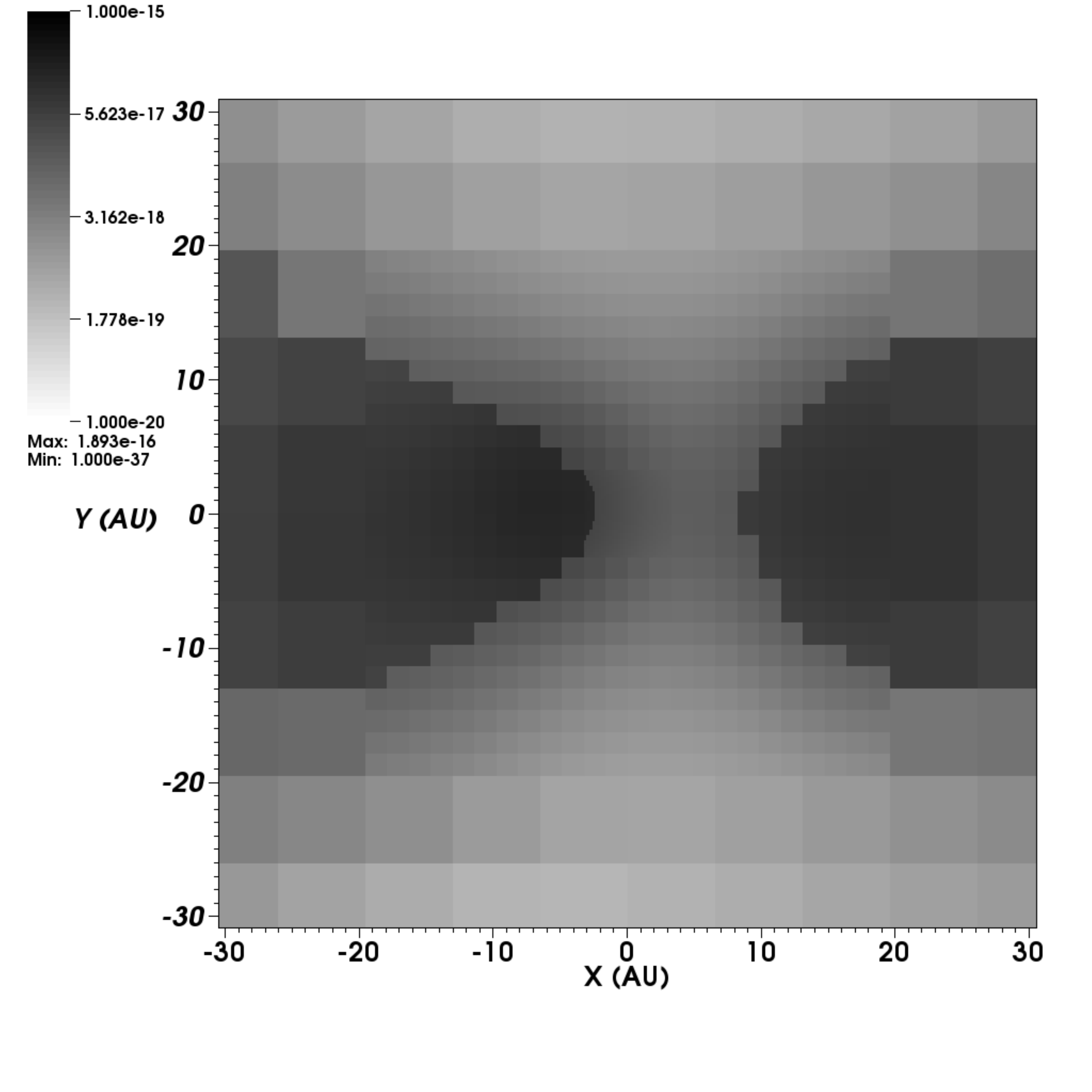}}
  \subfigure[$10^6$ particles]{\includegraphics[scale=0.22]{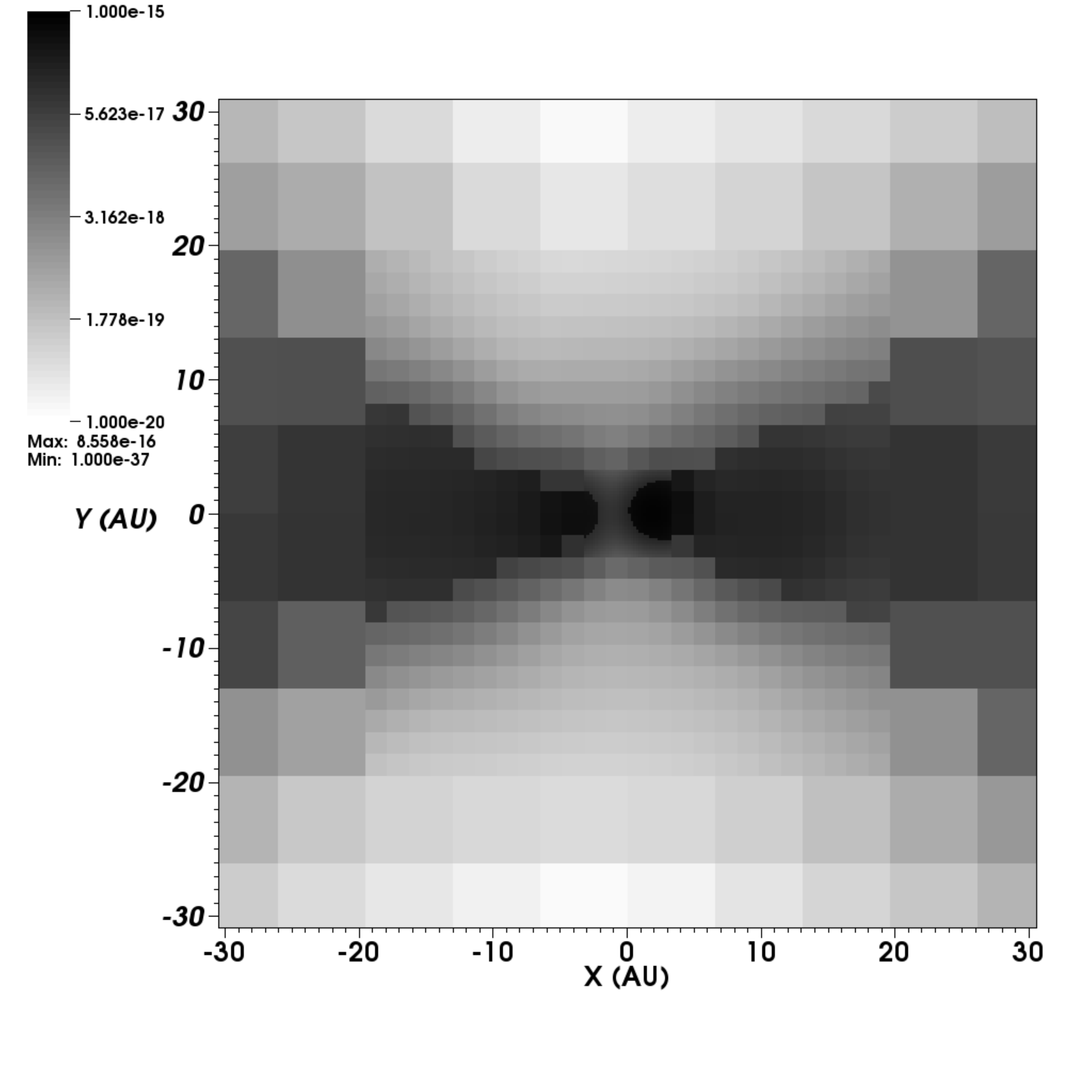}}
  \subfigure[$10^7$ particles]{\includegraphics[scale=0.22]{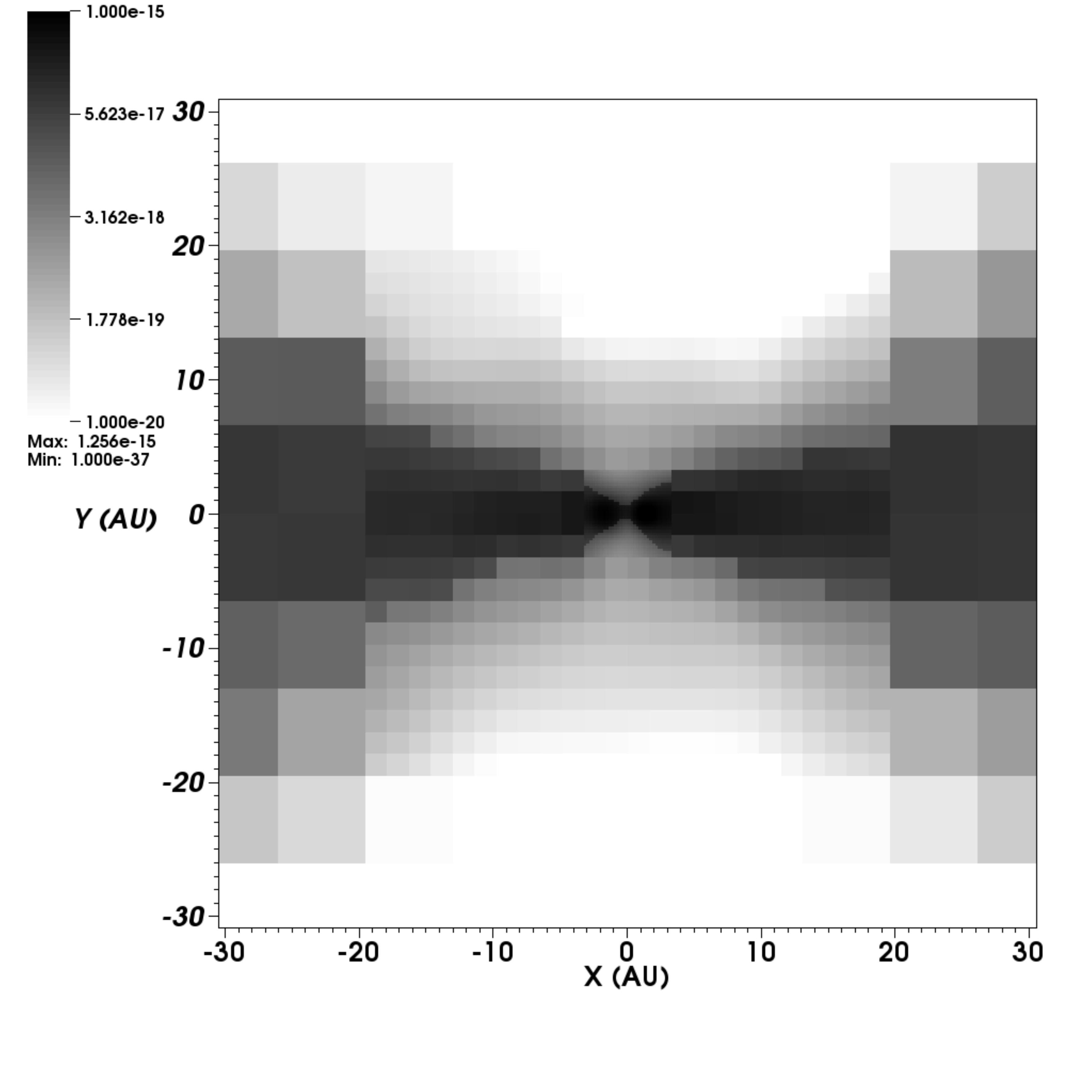}}
  \subfigure[$10^8$ particles]{\includegraphics[scale=0.22]{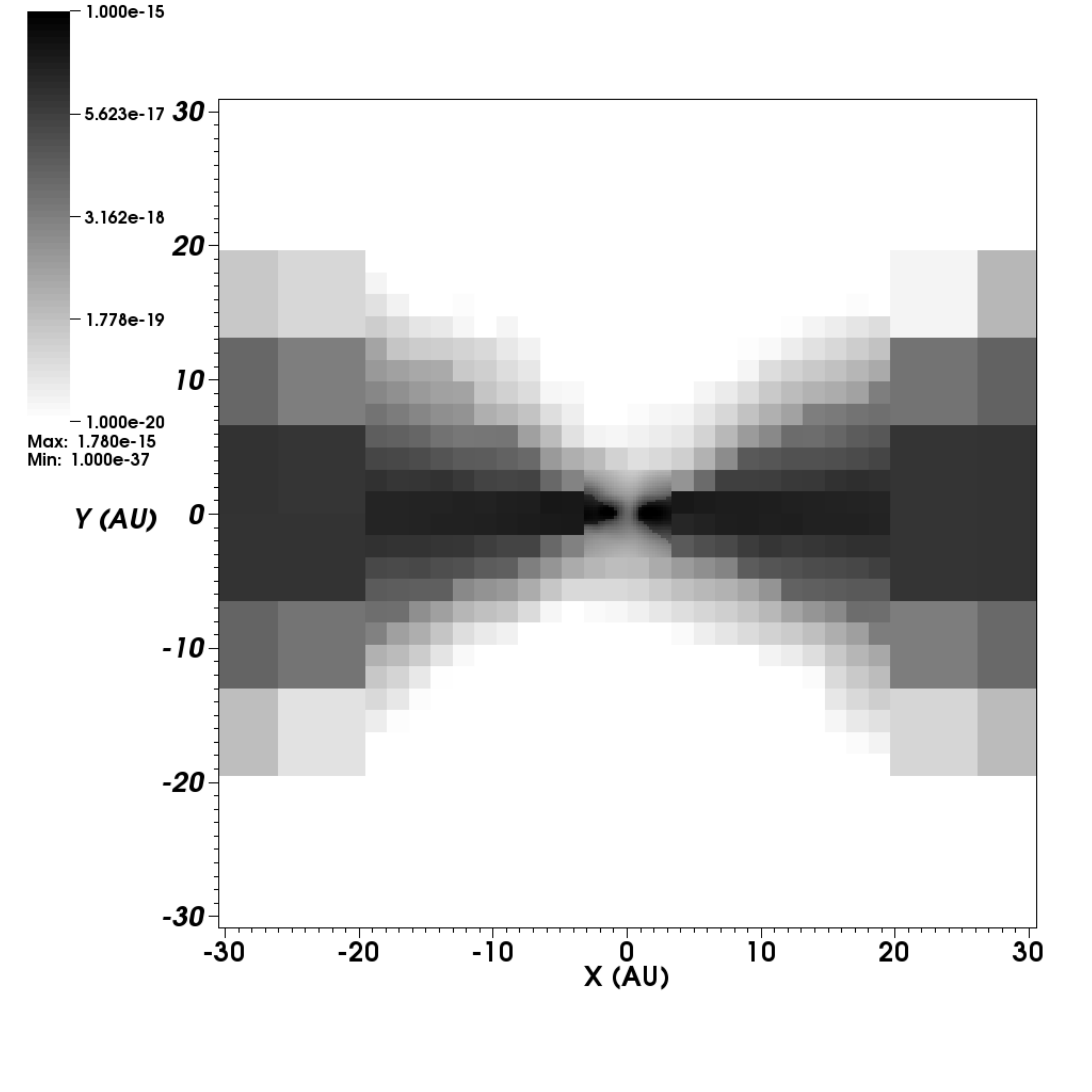}}
  \caption{Density ($\rm{g\,cm}^{-3}$)  in the central region of the disc as represented on AMR grids derived from different numbers
    of SPH particles. The AMR grid is generated using a mass per cell
    limit of $5\times10^{26}$\,g with extra resolution at the origin
    as described in Section~\ref{subsec:grid_res}.}
  \label{fig:n_part_central_dens}
\end{figure*}
The effective resolution of the AMR mesh is dependent on two factors; the probability of having sufficient particles to adequately sample the small volume of the disc inner-edge, and the particle smoothing length (which determines the effective spatial resolution of the SPH density representation).
With $10^5$ and $10^6$ particles a central gap is present but is too
large. With $10^7$ particles there is more material close in to the
source but the gap itself is not represented (because the smoothing length of the particles near the disc inner-edge is too large) and the source is
obscured. Only with $10^8$ particles is a gap of approximately the
correct size present.  The representation of the central 1\,au gap is
significantly affected by the number of particles used to represent
the disc, with a large number of particles required to achieve an
accurate density distribution in this spatially small region. 

Figure~\ref{fig:n_part_tem_diff} shows fractional errors in the
temperature distribution for different numbers of
particles. 
\begin{figure*}
  \subfigure[$10^5$ particles]{\includegraphics[scale=0.3]{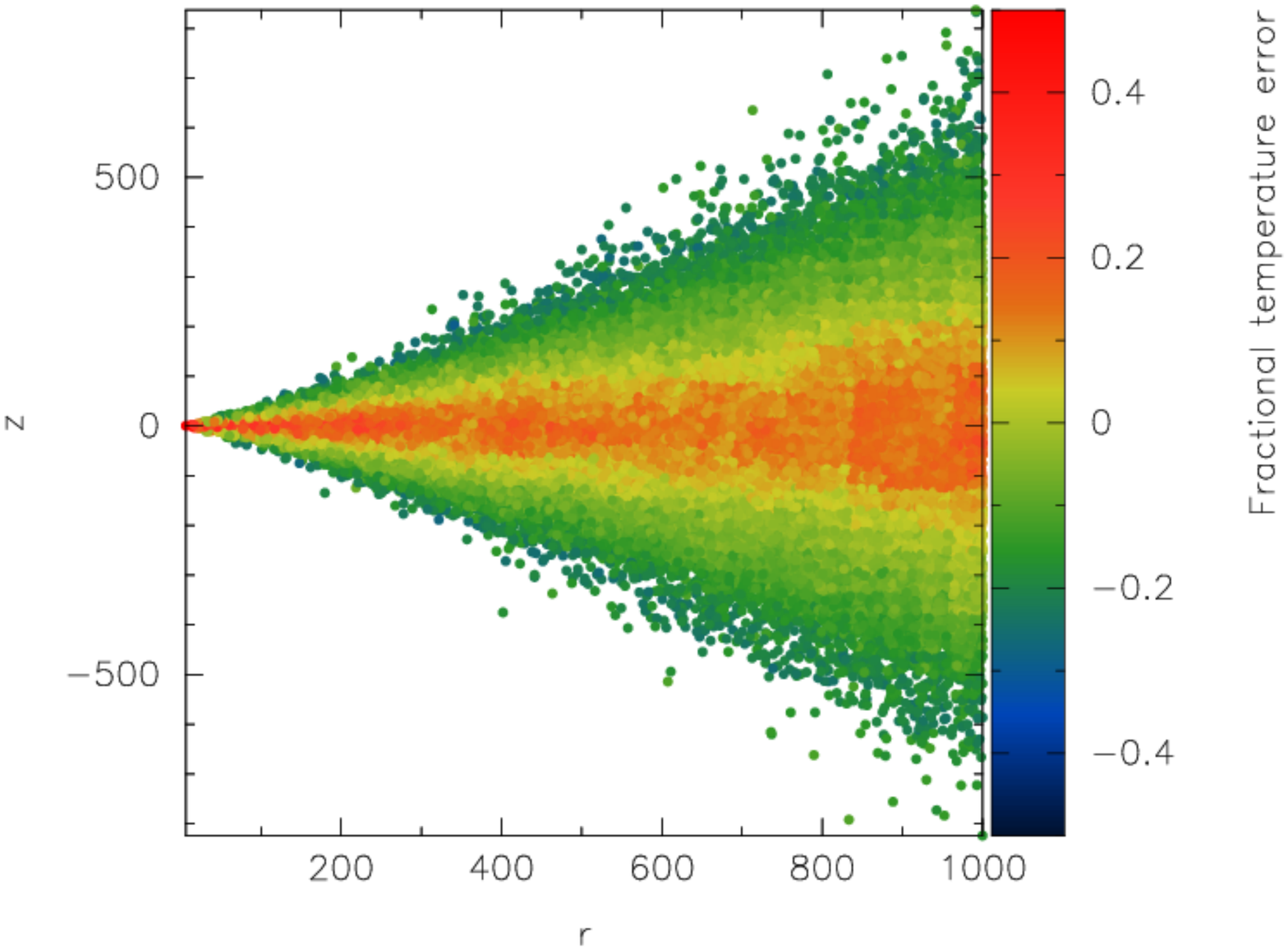}}
  \subfigure[$10^6$ particles]{\includegraphics[scale=0.3]{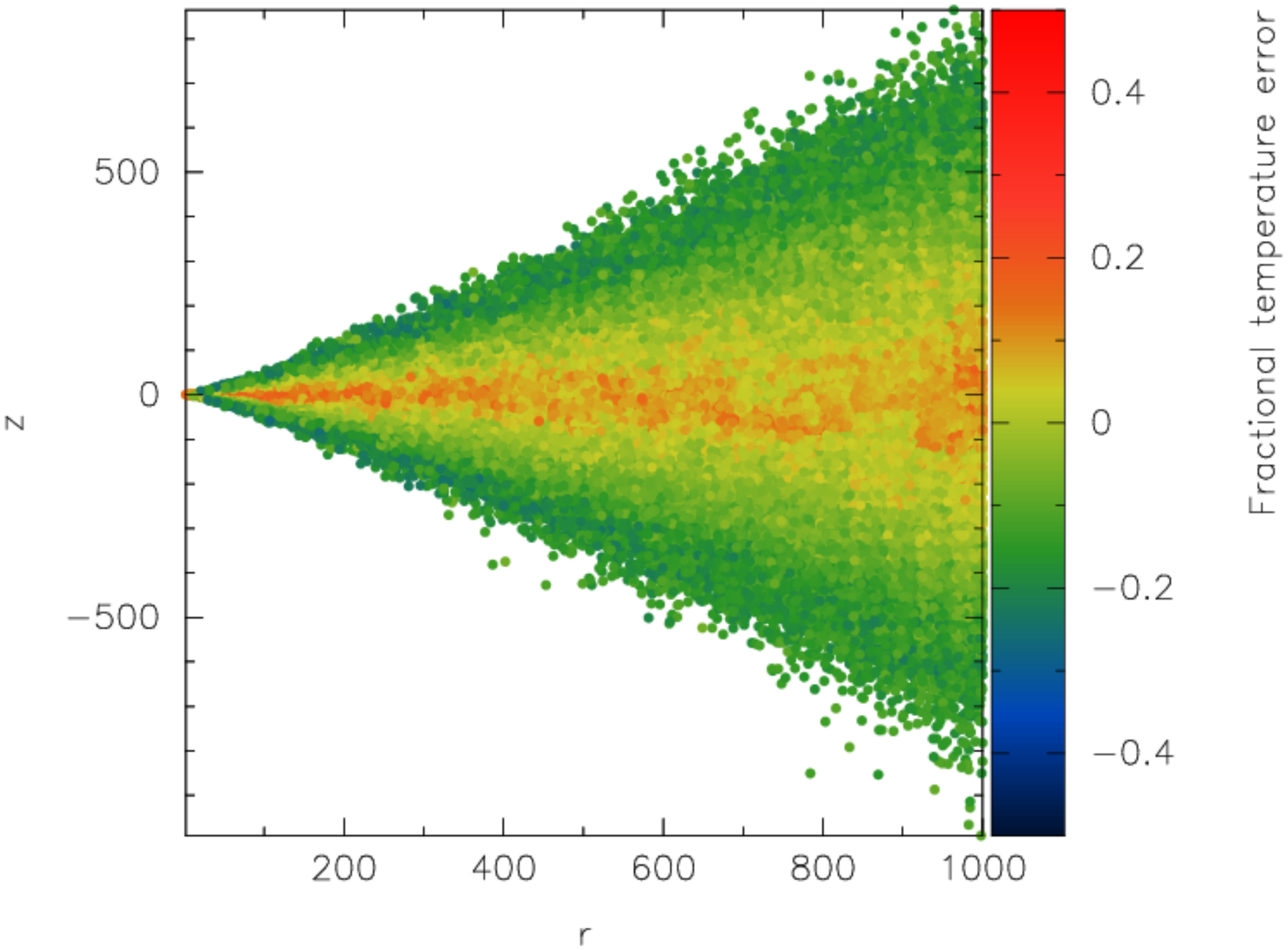}}
  \subfigure[$10^7$ particles]{\includegraphics[scale=0.3]{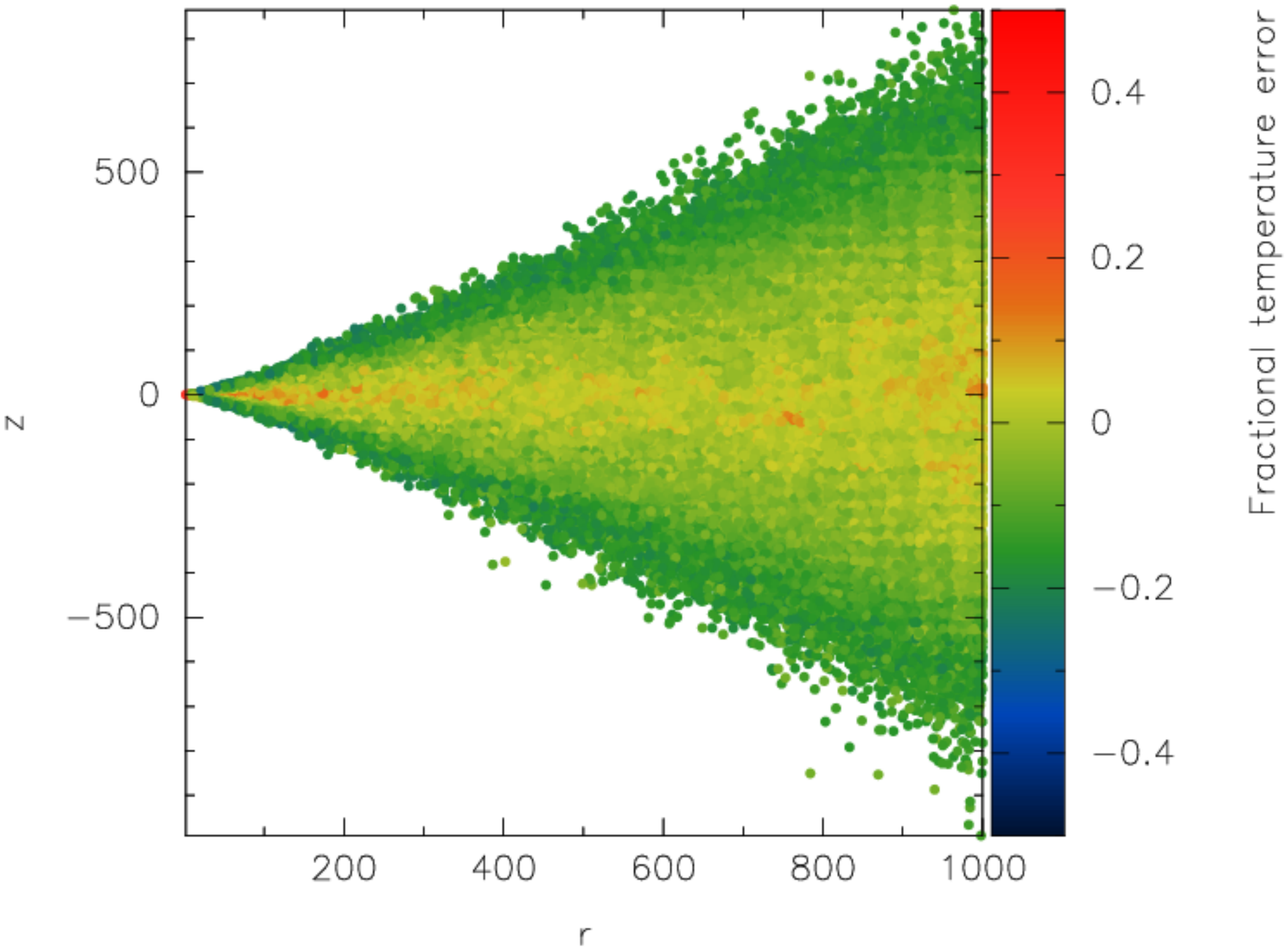}}
  \subfigure[$10^8$ particles]{\includegraphics[scale=0.3]{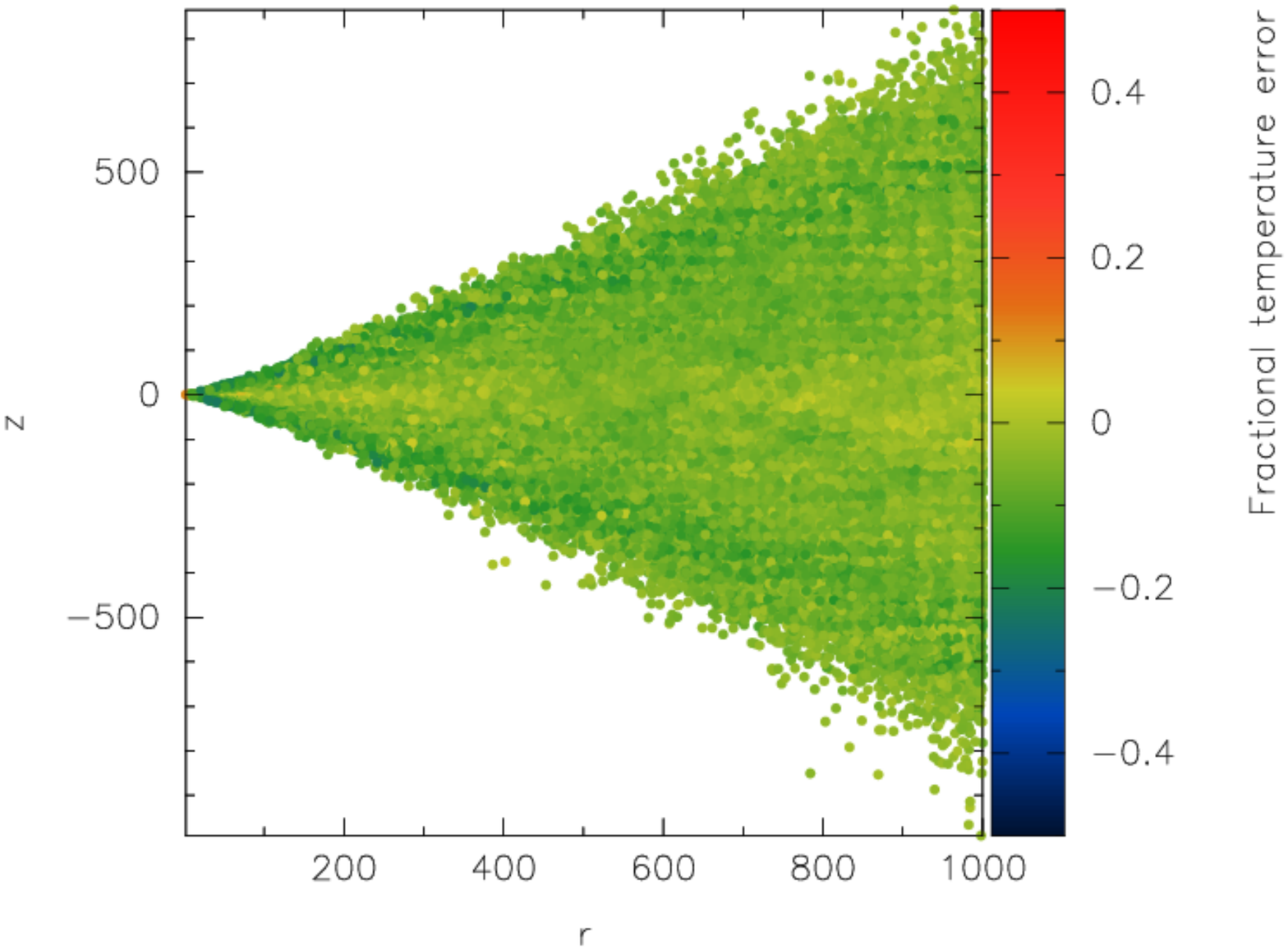}}
  \caption{Fractional errors in the temperature distribution of discs
    represented by $10^5$, $10^6$, $10^7$ and $10^8$ particles. The
    AMR grid is generated using a mass per cell limit of
    $5\times10^{26}$\,g with extra resolution at the origin as
    described in Section~\ref{subsec:grid_res}. The axis units are
    au. The corresponding density distributions in the
    central part of the disc are plotted in
    Fig.~\ref{fig:n_part_central_dens}.}
  \label{fig:n_part_tem_diff}
\end{figure*}
When using $10^5$ particles to represent the disc there is a too much
heating in the disc mid plane and insufficient heating further out of
the mid plane. The effect is reduced as the number of particles is
increased and reaches a very low level with $10^8$ particles, at which stage the
temperature distribution agrees with the benchmark to within $\sim$10\%.  

To allow a more quantitative comparison the mid-plane temperature and
the benchmark mid-plane temperature are plotted in
Fig.~\ref{fig:n_part_tem_diff_midpane} for all particles within 0.01
scale heights of the mid-plane. The fractional temperature error for
each particle is also plotted where a positive error indicates a
temperature hotter than the benchmark value. 
\begin{figure*}
  \subfigure[$10^5$ particles]{\includegraphics[scale=0.3]{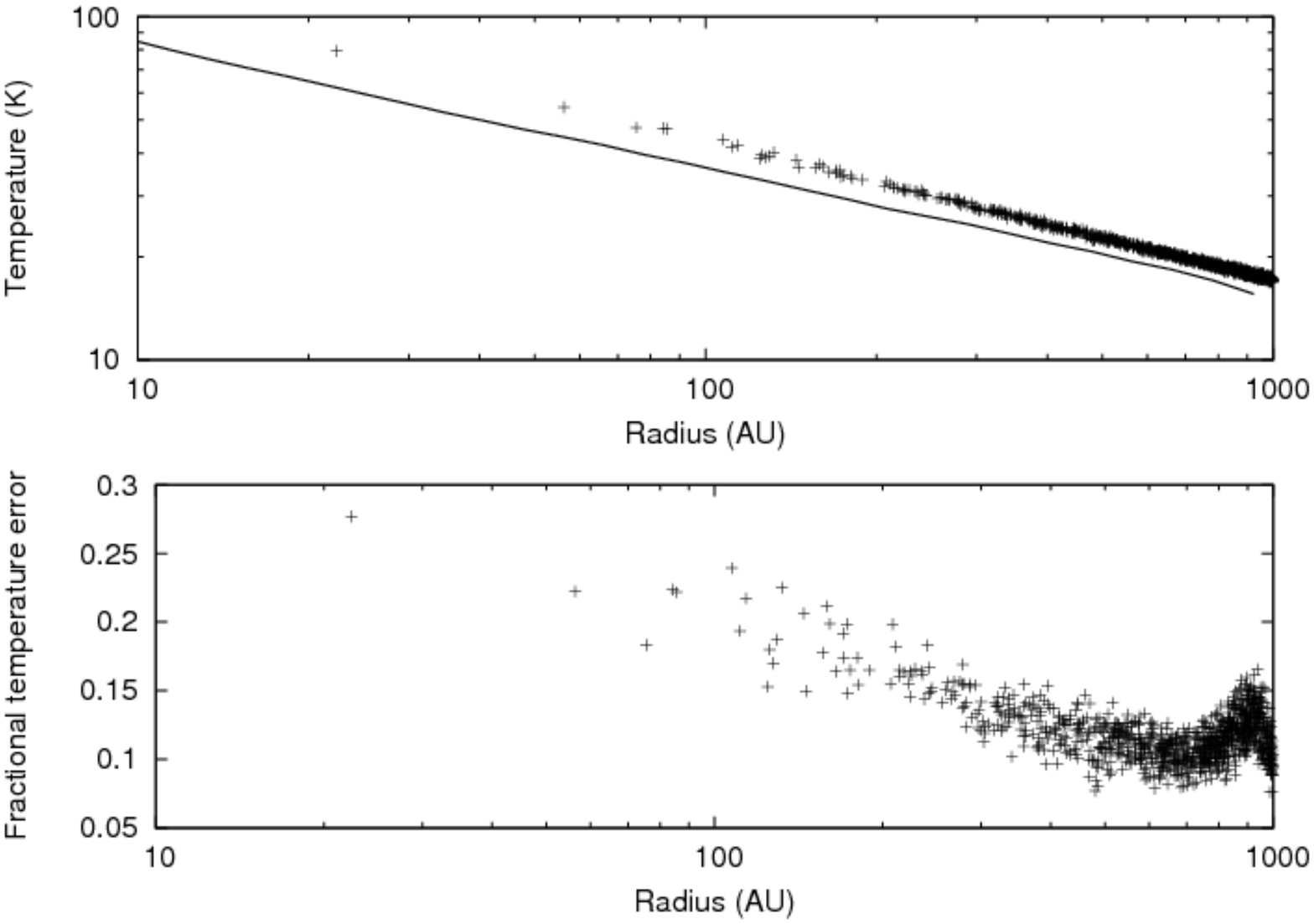}}
  \subfigure[$10^6$ particles]{\includegraphics[scale=0.3]{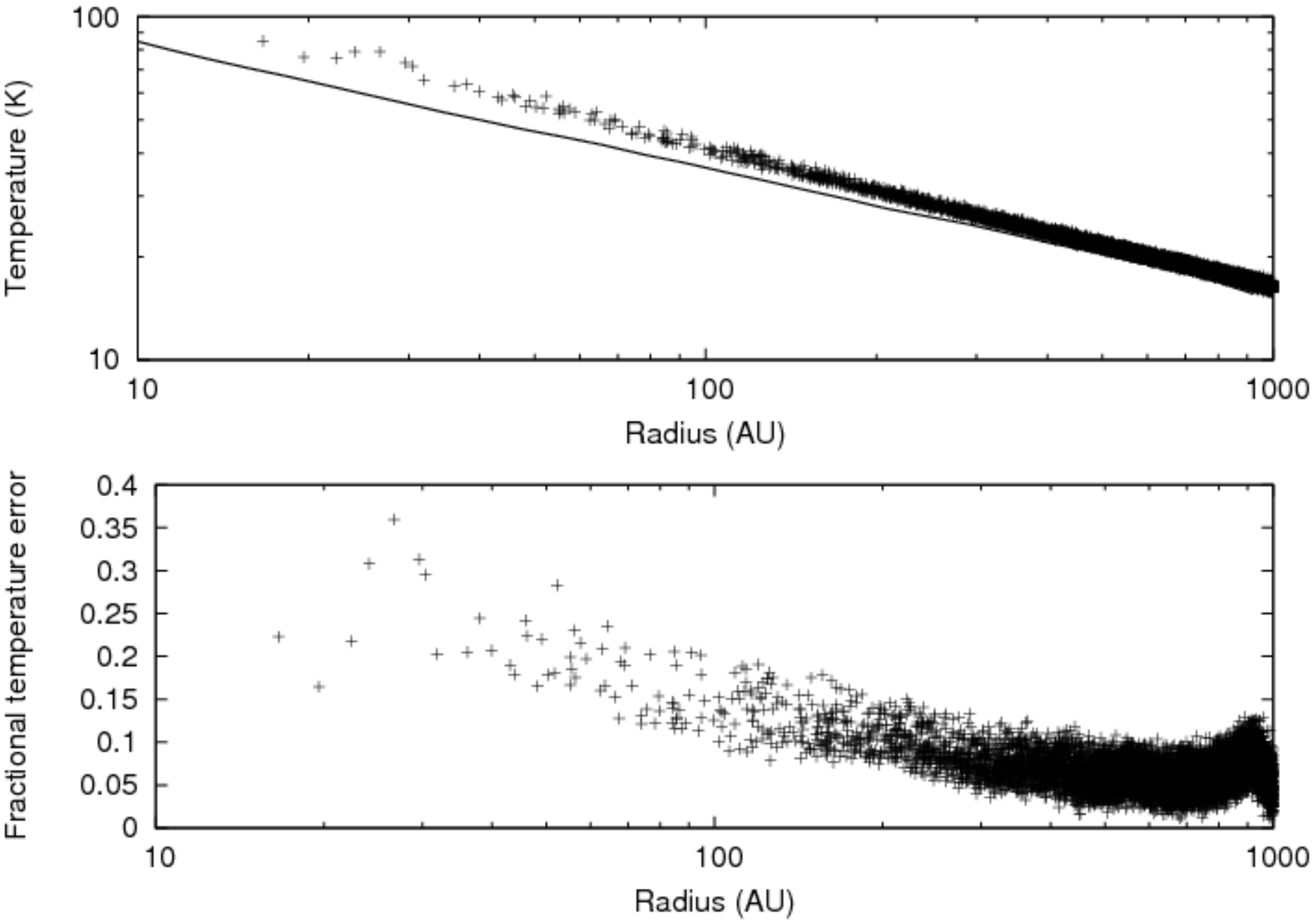}}
  \subfigure[$10^7$ particles]{\includegraphics[scale=0.3]{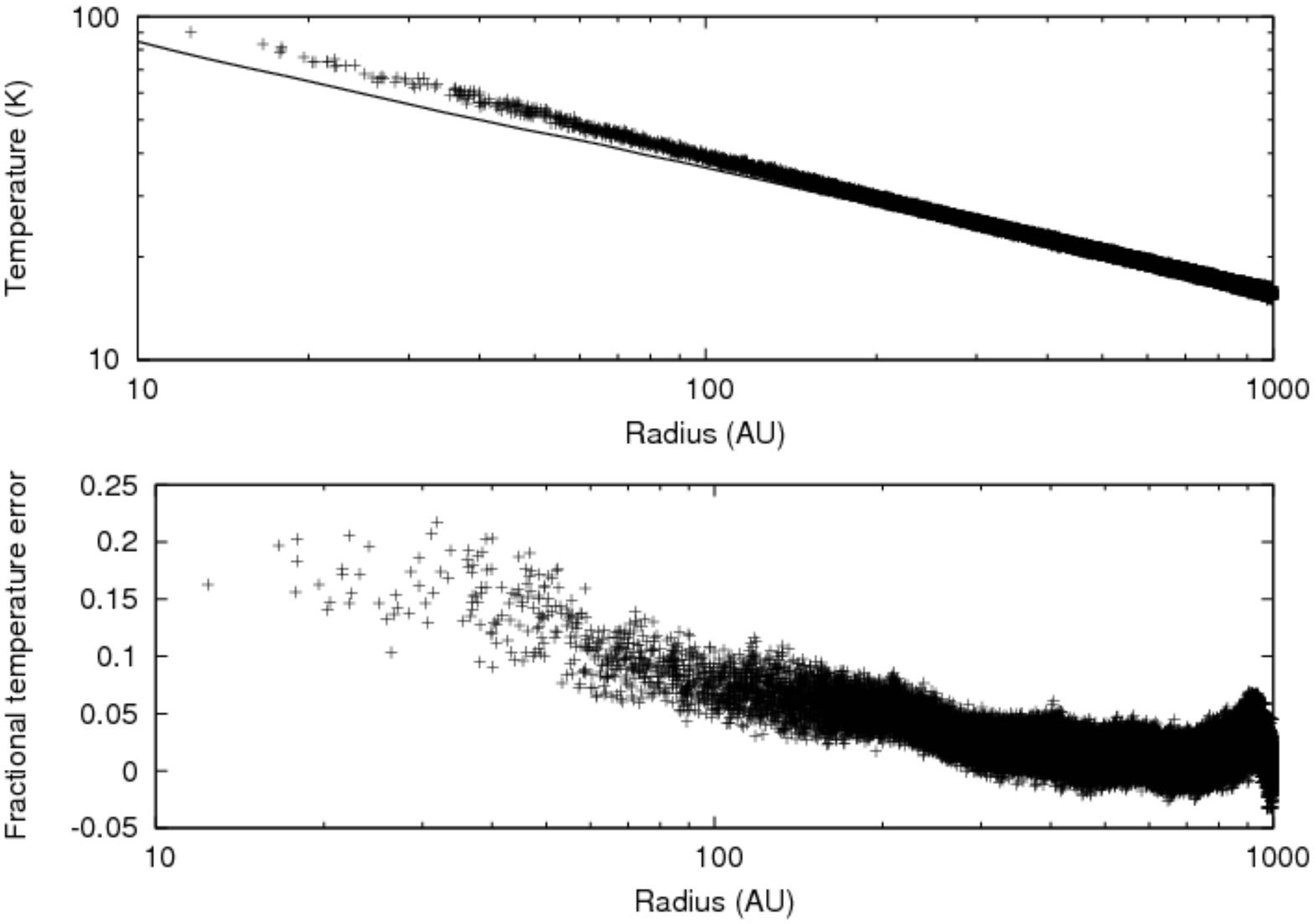}}
  \subfigure[$10^8$ particles]{\includegraphics[scale=0.3]{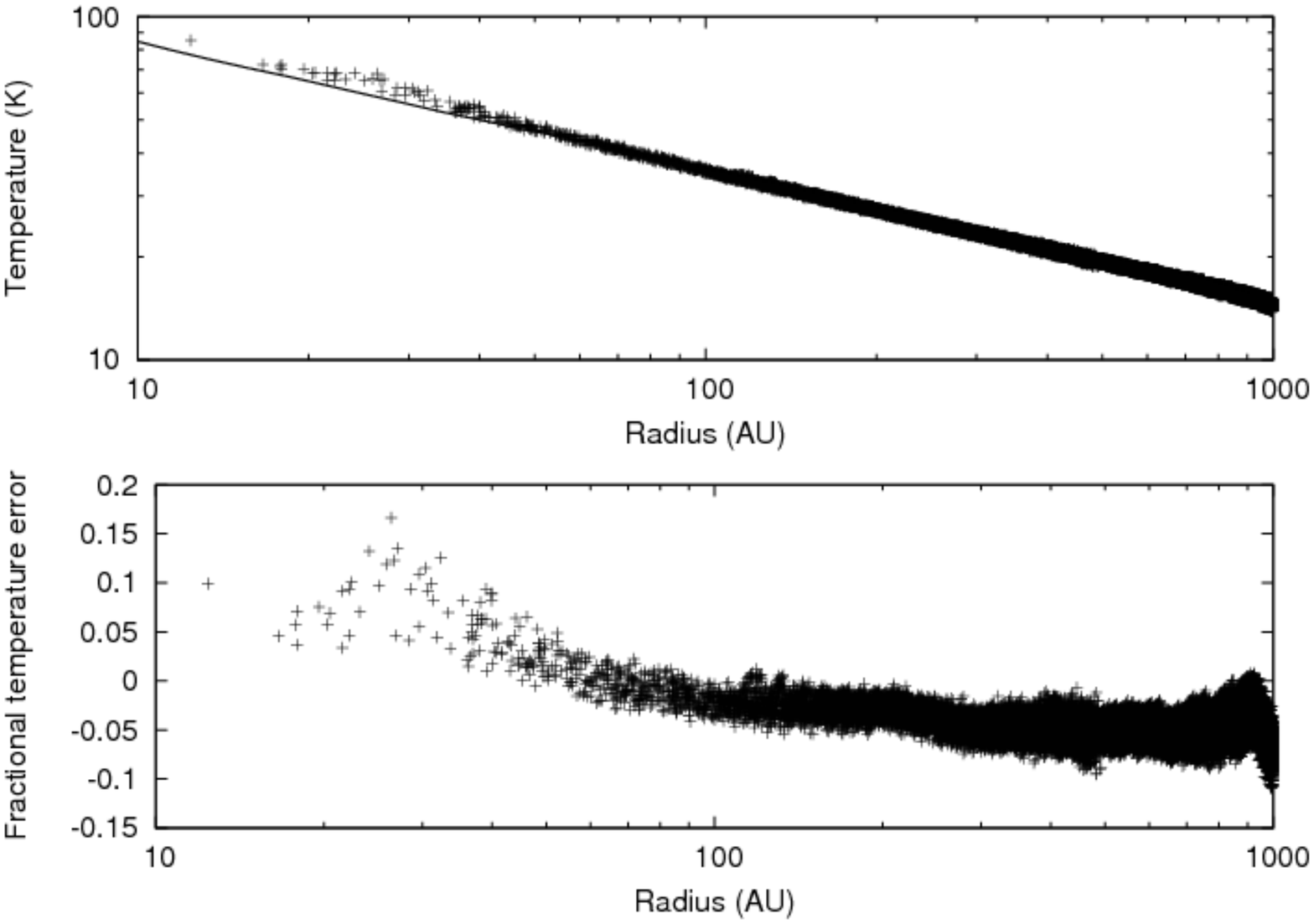}}
  \caption{Fractional temperature error, as a function of radius, for particles within 0.01
    scale heights of the mid-plane for discs represented by $10^5$,
    $10^6$, $10^7$ and $10^8$ particles. }
  \label{fig:n_part_tem_diff_midpane}
\end{figure*}

SEDs are plotted in Fig.~\ref{fig:n_part_sed_013} for a viewing angle
of 12.5 degrees and Fig.~\ref{fig:n_part_sed_077} for a viewing angle
of 77.5 degrees. At the shortest wavelengths the 12.5 degrees
inclination angle SED is dominated by the photosphere of the central
star. This region of the SED is well represented with $10^5$ and
$10^8$ particles but is of course less well represented with $10^6$ or
$10^7$ particles where the central star is obscured. At the longest
wavelengths the disc is optically thin and the SEDs will be well
modelled provided the total disc mass is accurately represented. In
all four cases the total disc mass is well represented (see
Section~\ref{section:grid_generation}) and the long wavelength SED
matches the benchmark results. At intermediate wavelengths the SED is
sensitive to the details of the disc representation and it is clear
that accurately modelling the SED from this disc is challenging, even
with $10^8$ particles. 
At an inclination angle of 77.5 degrees the central star is
    overly obscured by disc material.  If the spatial resolution at the inner
    edge is low then the inner edge will be too thick (in the z
    direction) and too close to the central star. As a result there
    will be overly dense material in the observer's line of
    sight. An SED with insufficient emission from the central star at
    this inclination angle is indicative that the inner edge is not
    well resolved.

\begin{figure*}
  \subfigure[$10^5$ particles]{\includegraphics[scale=0.3]{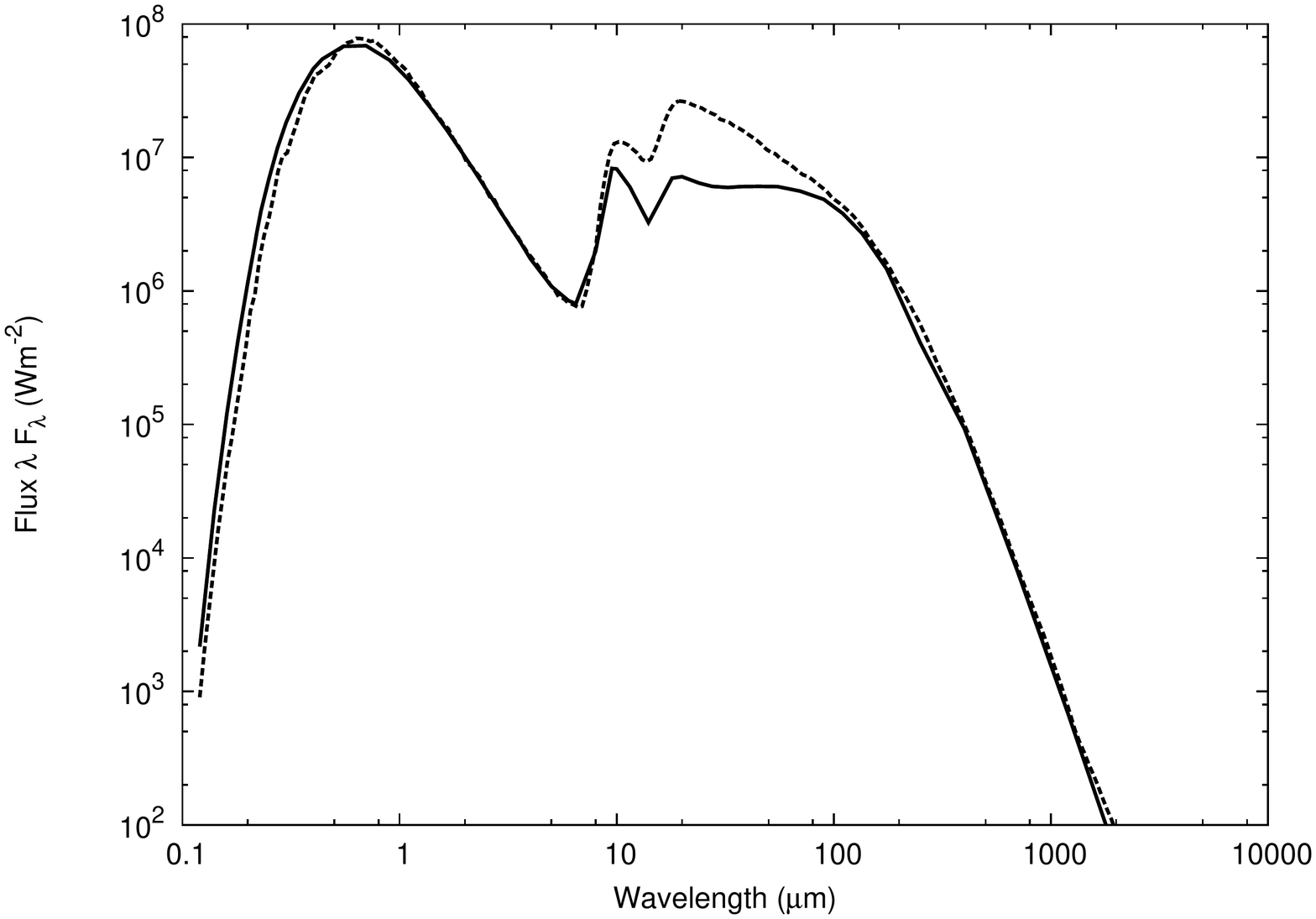}}
  \subfigure[$10^6$ particles]{\includegraphics[scale=0.3]{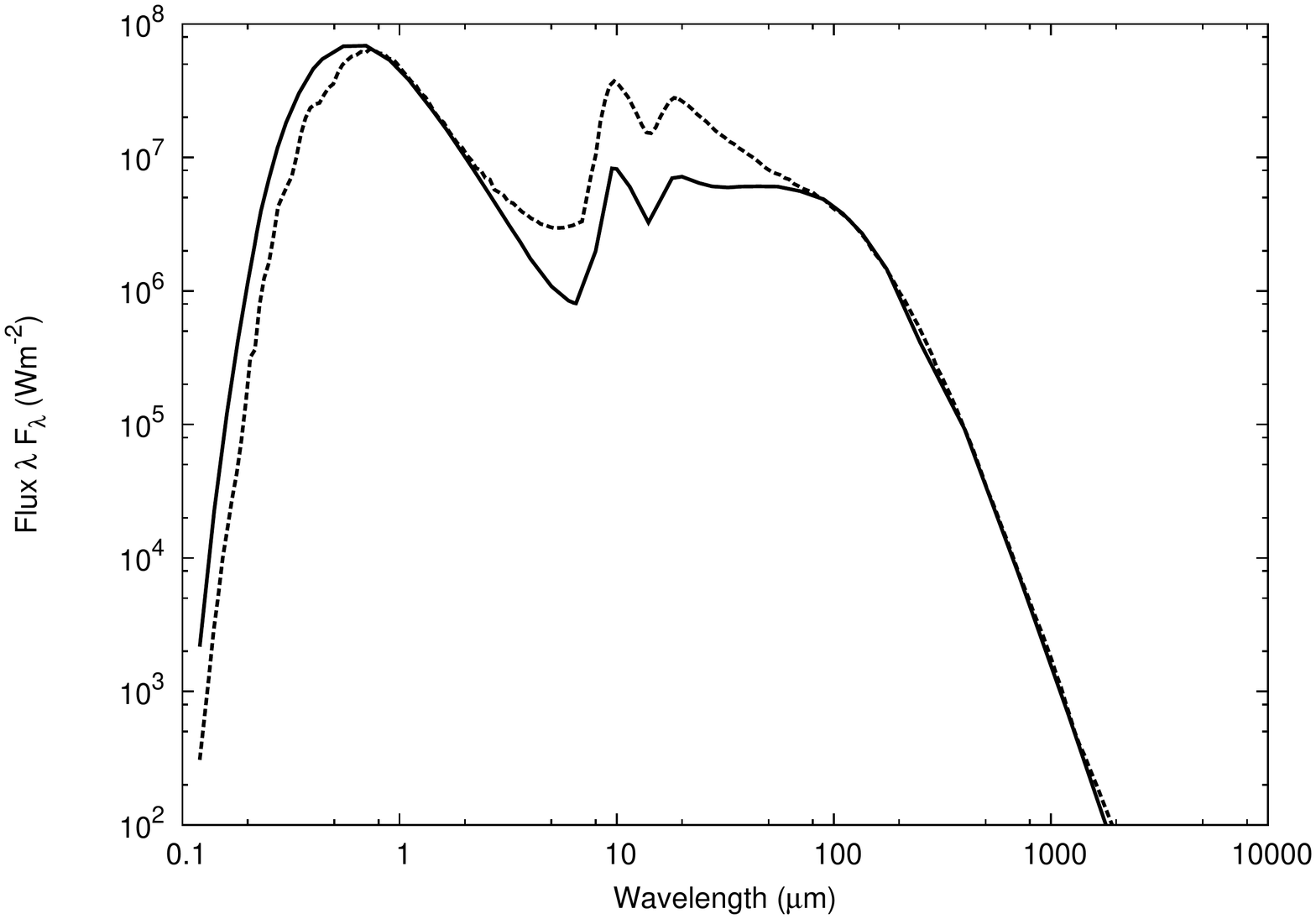}}
  \subfigure[$10^7$ particles]{\includegraphics[scale=0.3]{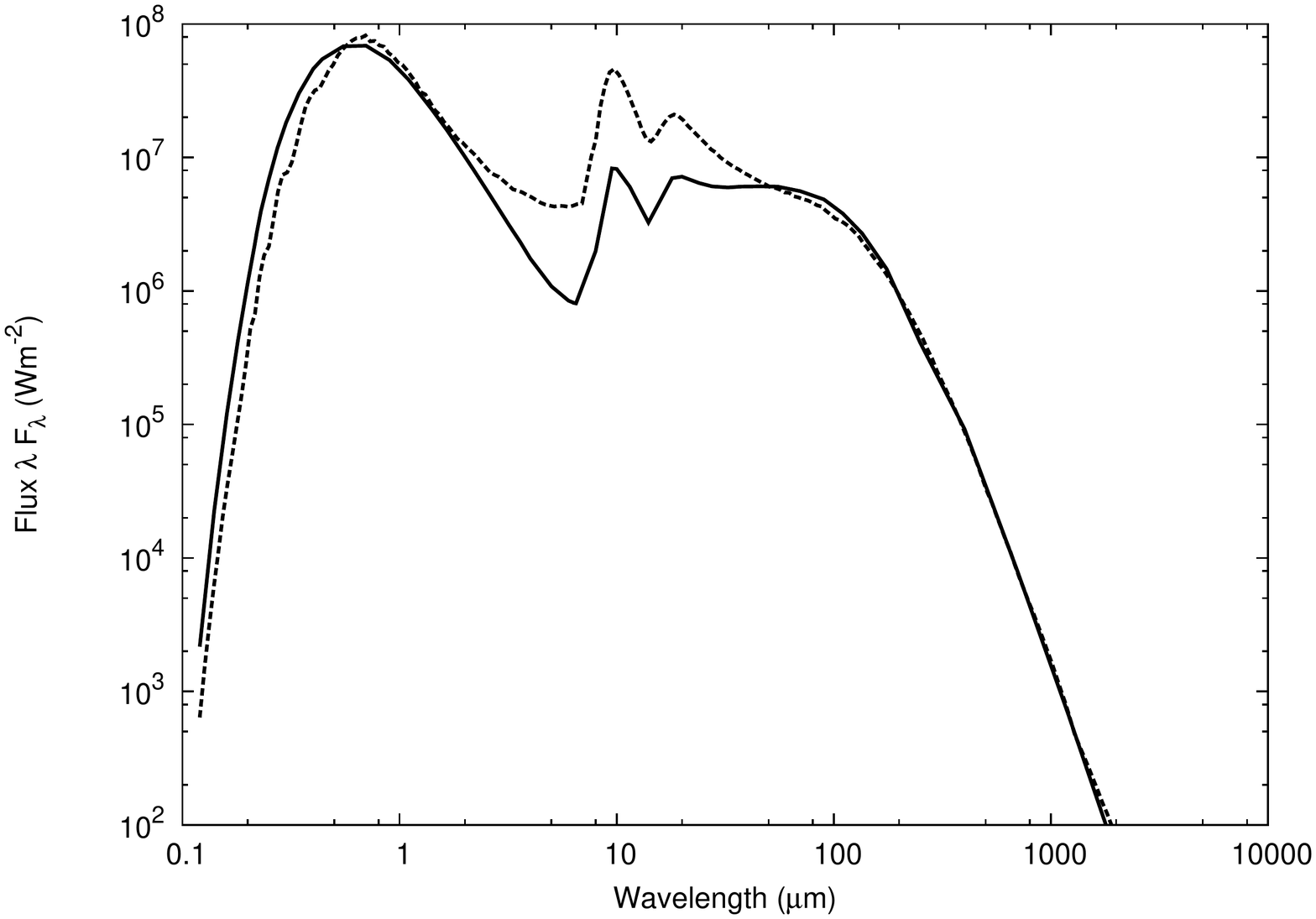}}
  \subfigure[$10^8$ particles]{\includegraphics[scale=0.3]{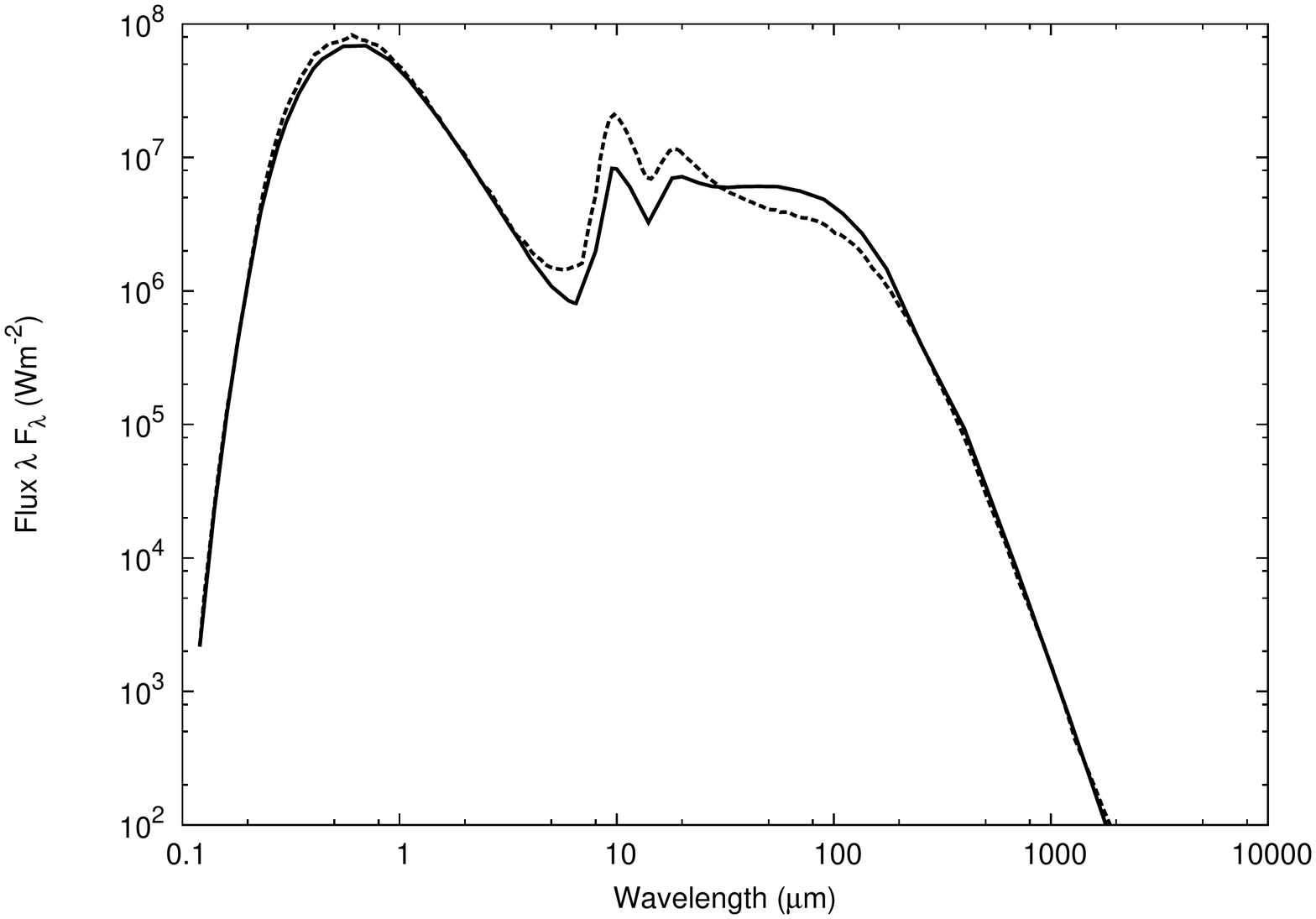}}
  \caption{SEDs at a 12.5 degree inclination angle for discs
    represented by $10^5$, $10^6$, $10^7$ and $10^8$ particles. The
    benchmark result is plotted as a solid line and the {\sc torus}
    SEDs are plotted as dashed lines. The corresponding density
    distributions in the central part of the disc are plotted in
    Fig.~\ref{fig:n_part_central_dens} and fractional temperature
    errors are plotted in Fig.~\ref{fig:n_part_tem_diff}.}
  \label{fig:n_part_sed_013}
\end{figure*}
\begin{figure*}
  \subfigure[$10^5$ particles]{\includegraphics[scale=0.3]{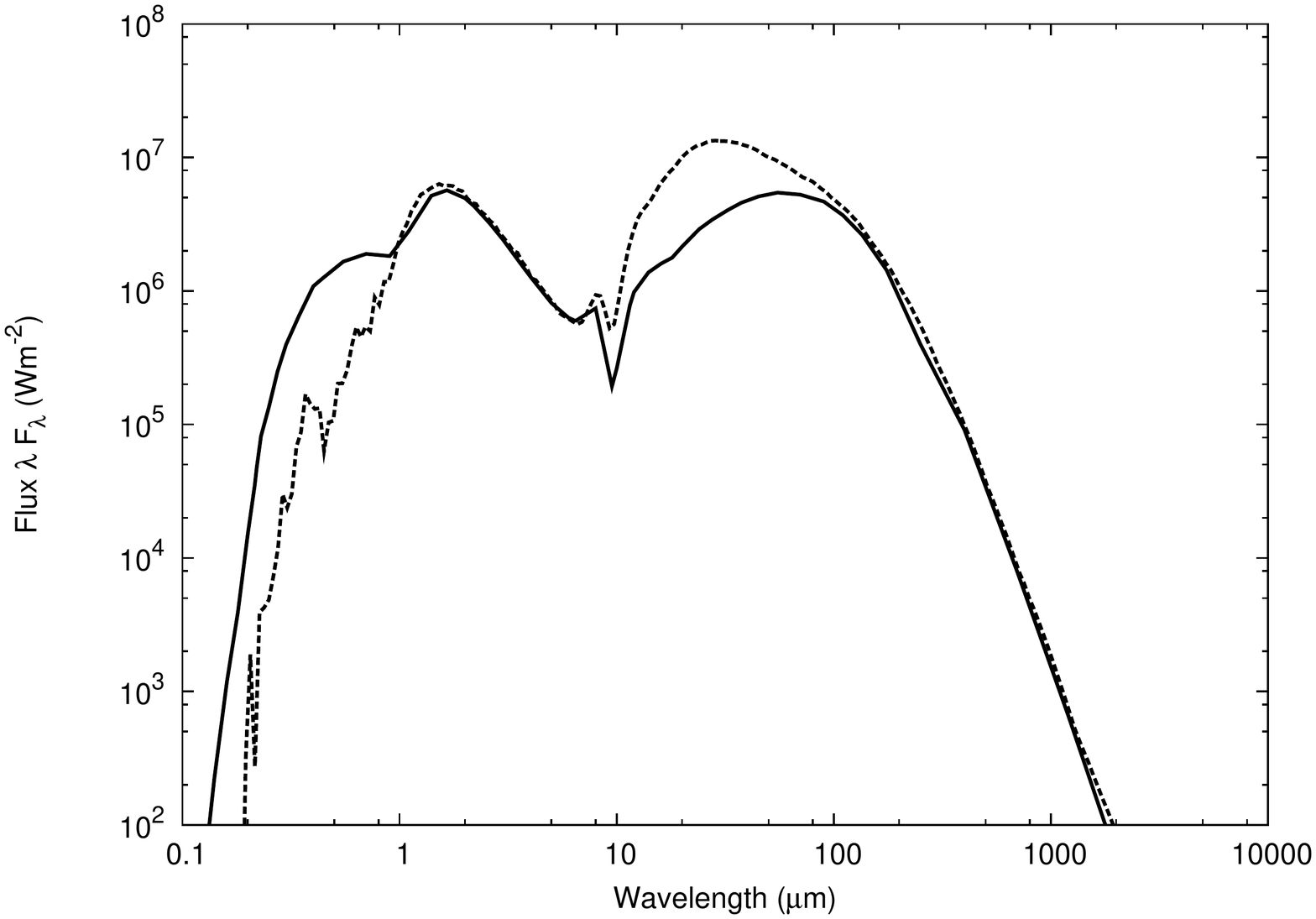}}
  \subfigure[$10^6$ particles]{\includegraphics[scale=0.3]{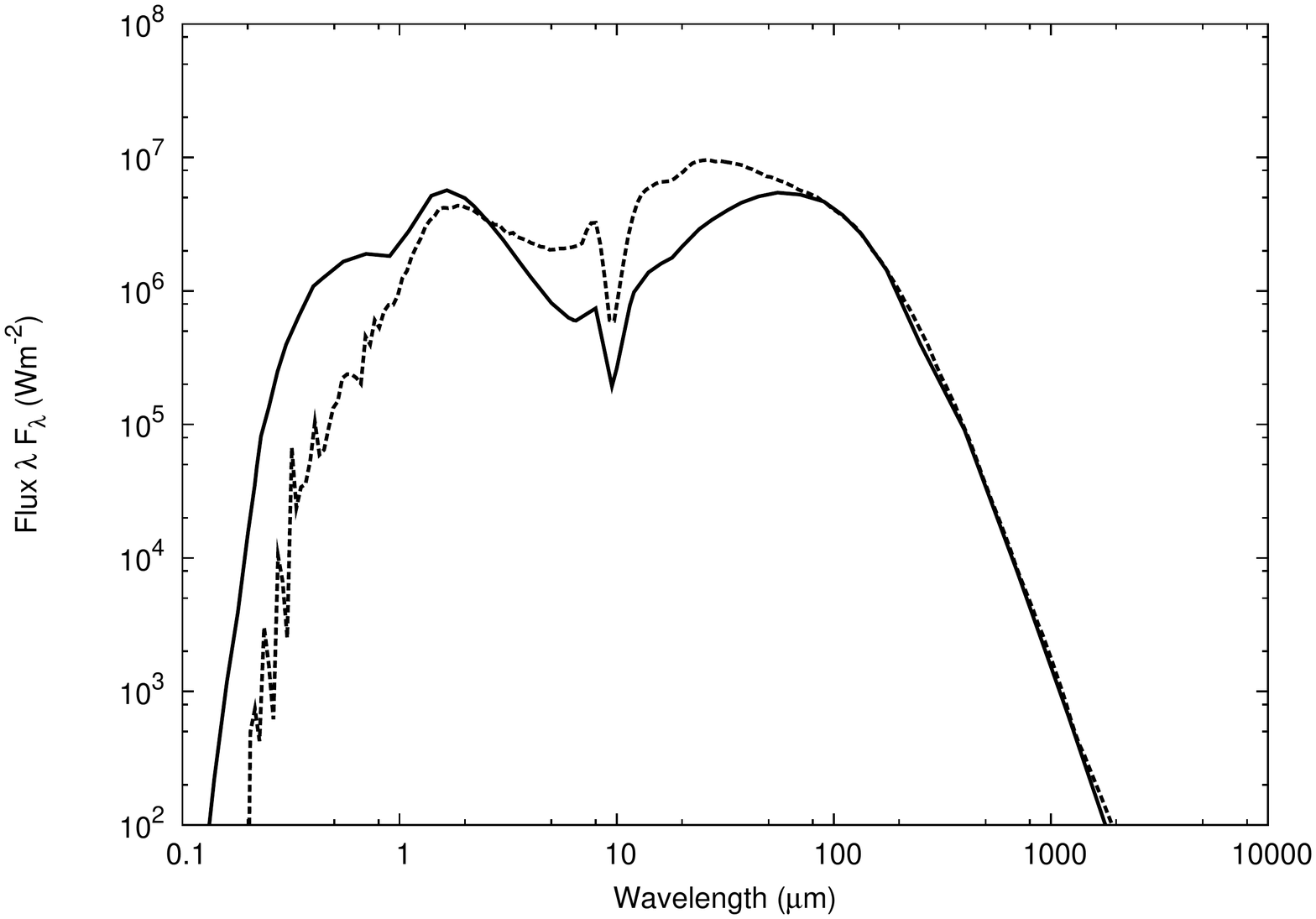}}
  \subfigure[$10^7$ particles]{\includegraphics[scale=0.3]{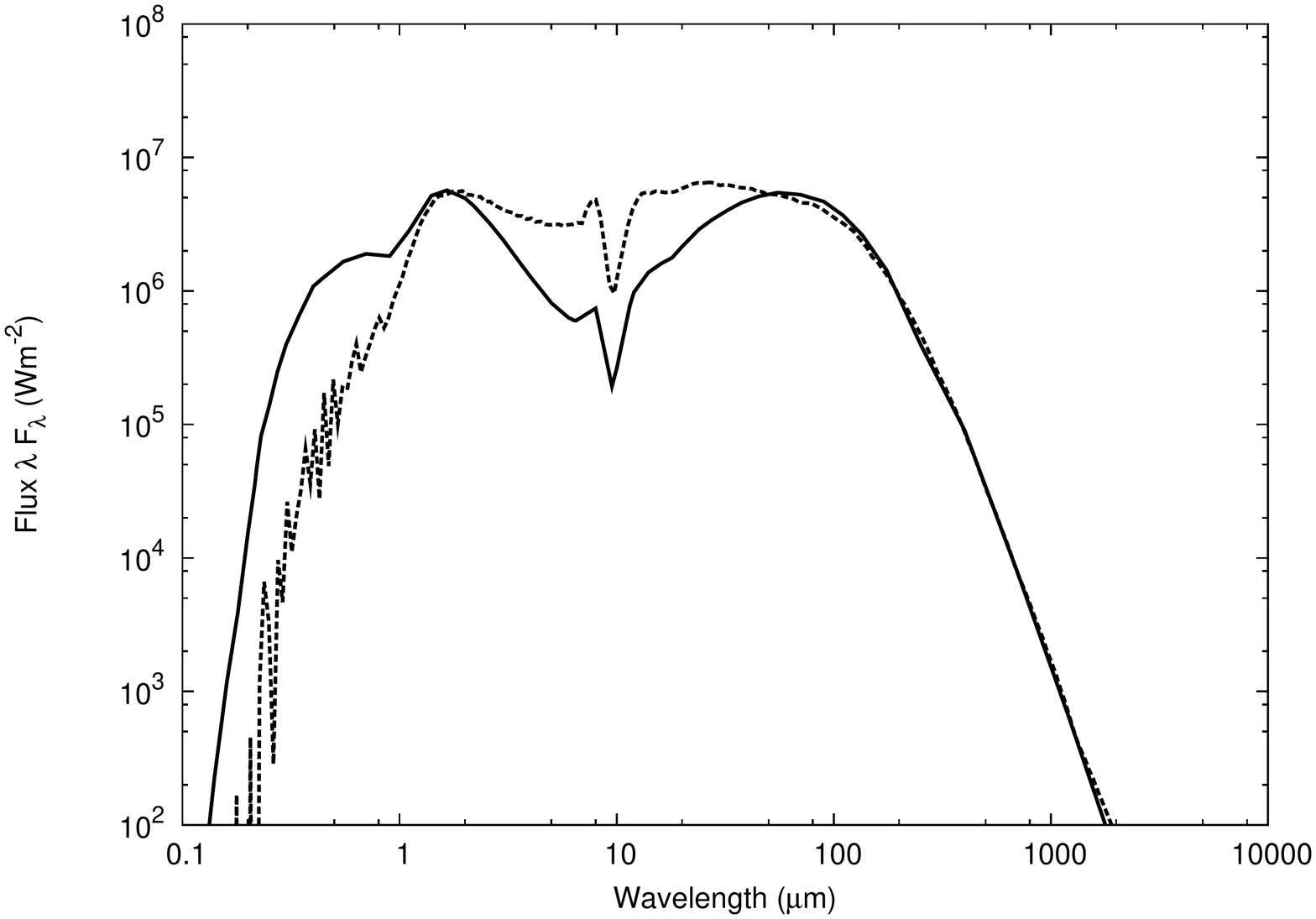}}
  \subfigure[$10^8$ particles]{\includegraphics[scale=0.3]{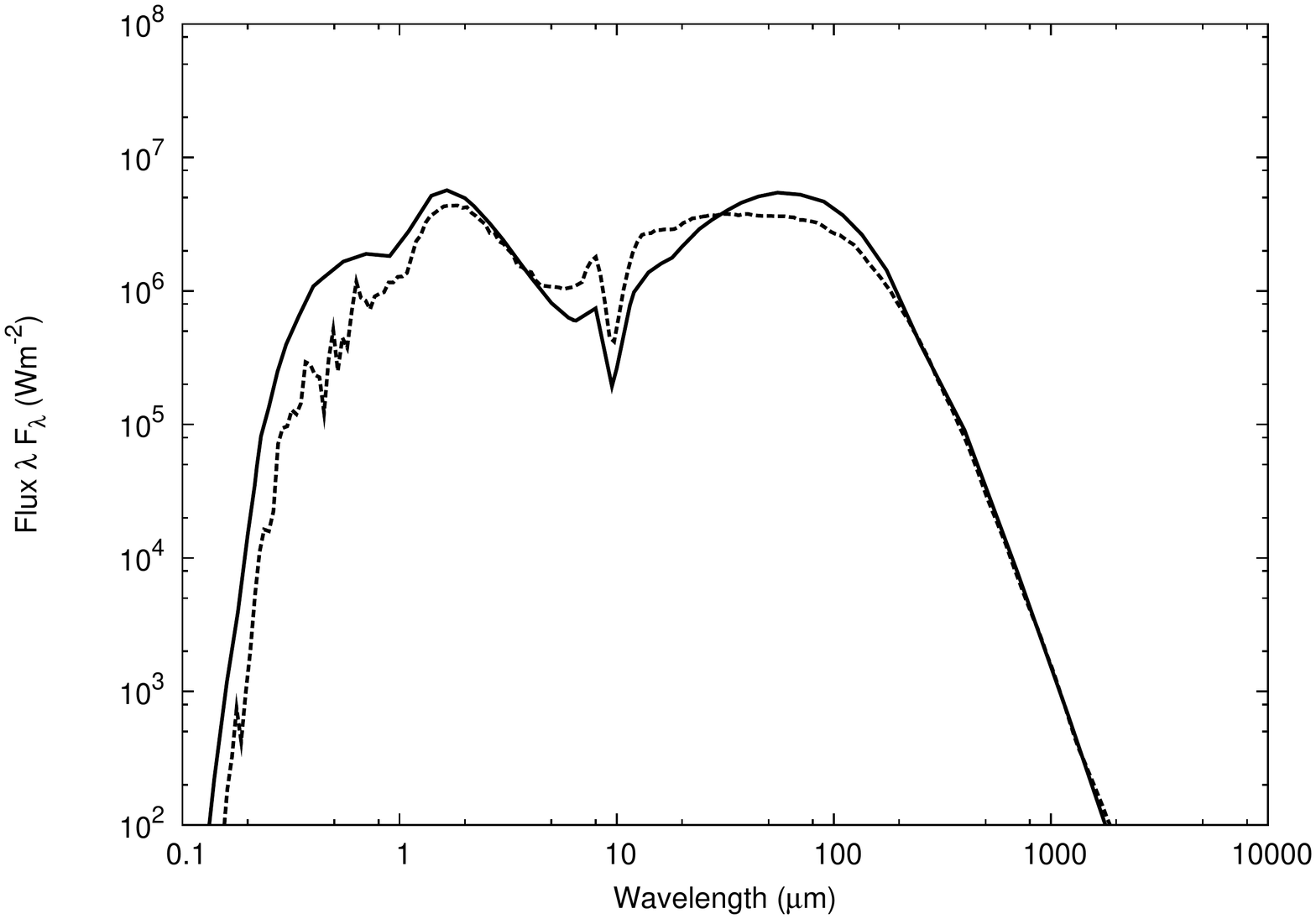}}
  \caption{SEDs at a 77.5 degree inclination angle for discs
    represented by $10^5$, $10^6$, $10^7$ and $10^8$ particles. The
    benchmark result is plotted as a solid line and the {\sc torus}
    SEDs are plotted as dashed lines. The corresponding density
    distributions in the central part of the disc are plotted in
    Fig.~\ref{fig:n_part_central_dens} and fractional temperature
    errors are plotted in Fig.~\ref{fig:n_part_tem_diff}.}
  \label{fig:n_part_sed_077}
\end{figure*}

The results of our benchmark tests have shown that the temperature
distribution within the disc can be calculated with a typical accuracy
of better than 20 per cent with only $10^6$ particles (see
Fig.~\ref{fig:n_part_tem_diff} and
Fig.~\ref{fig:n_part_tem_diff_midpane}), which offers a significant
improvement over the use temperatures derived from an eos. Although
relatively few particles are required to give a temperature
distribution sufficiently accurate for hydrodynamic calculations,
obtaining an accurate SED from the model disc would require many more
particles, as there are significant discrepancies relative to the
benchmark even with $10^8$ particles (see
Fig.~\ref{fig:n_part_sed_013} and Fig.~\ref{fig:n_part_sed_077}).

\section{Coupled SPH and Monte-Carlo radiative transfer simulation of
  a circumstellar disc}

\label{section:sph_disc}

The above analysis indicates that the hybrid code is capable of
successfully passing the density distribution from the hydrodynamics
step to the radiative-transfer code, and that 
temperatures sufficiently accurate for hydrodynamic calculations
    (under the assumption of radiative equilibrium)
can be passed back to the SPH code in
order to update the gas pressure and perform the next hydrodynamical
step. In order to test the feasibility of performing a
radiation-hydrodynamics calculation using the hybrid code we adopt a
`realistic' test problem, that of a circumstellar disc around a
classical T Tauri star. The mid-plane optical depths here are higher
than that of the Pascucci benchmark, providing a more demanding
challenge both in terms of computation time and required numerical
accuracy.

\subsection{Description of model}

A model system was constructed similar to the classical T Tauri star
AA Tau. \cite{o'sullivan_2005} determine the properties of AA Tau
using Monte-Carlo modelling of the observed SED, and we use these
derived properties as a basis for constructing our model system. The
disc of $0.02\,{M}_{\odot}$ is initialised with an outer radius of
150\,au, and an inner radius of 1.0\,au. 
The inner radius of our model disc is enlarged, compared to the
    inner radius determined by \cite{o'sullivan_2005}, as resolving a
    smaller inner edge in a disc of this size would require many more
    SPH particles. The inner edge of our model disc will be cooler
    than the inner edge of a disc with a smaller radius but the
    important physical processes are captured (i.e. there is an optically
    thick inner edge which can puff up and shadow the disc causing a
    hydrodynamic response.)
The central star has a mass of mass $1\,M_{\odot}$ and is represented
by a sink particle which accretes gas particles within an accretion
radius of 1.0\,au. The disc itself is initially represented using
$10^6$ particles of equal mass, although some of these particles are
accreted by the central sink particle as the simulation
progresses. The central star has a radius of $2\,\rm{R}_{\odot}$, a
temperature of 4000\,K and a luminosity of $0.9\,\rm{L}_{\odot}$. 
The source spectrum taken to be a black body.

The disc is initially evolved using only the SPH part of the code in
order to bring the disc into dynamical equilibrium. 
The SPH implementation is that described by \cite{price_2007b}
    which uses the variable smoothing length formulation given in
    eqn~\ref{eqn:smlen}. Density and smoothing length are calculated
    iteratively according to the method of \cite{price_2007c} with
    approximately 60
    neighbours on average. An artificial viscosity term is calculated using the scheme described by
\cite{monaghan_1992} with parameters $\alpha=1$ and $\beta=2$.
During this initial phase the internal energies of the SPH particles are
determined using an equation of state which is isothermal
perpendicular to the plane of the disc and has a $T\propto 1/r$ radial
dependence. The disc is evolved for approximately 4000 years of
simulated time in order to allow transient effects from the
initialisation to dissipate. As the disc settles towards an
equilibrium state there are vertical motions which cause fluctuations
in the envelope of the disc, with the inner regions of the disc
reaching equilibrium more rapidly than the outer regions.

The disc is then evolved using the combined SPH and radiative transfer
code. In this configuration the internal energies of the SPH particles
are set using the temperatures calculated by the radiative transfer
code. The SPH part of the code does not change the particle
temperatures and they retain the temperature assigned by the radiative
transfer code until the next radiation time step. The temperatures
calculated by the radiative transfer code are equilibrium
temperatures, as it is assumed that the time scale for radiative
equilibrium is short compared to the dynamical time scale of the disc
(i.e. the ``prompt escape'' regime described by
\citealt{nayakshin_2009}).  
Calculating temperatures from the
    radiative transfer scheme alone is only a valid approximation
    provided heating from stellar radiation dominates other sources of
    heating (e.g. viscous or shock heating). The addition of a time
    dependent radiative transfer scheme would allow for the inclusion
    of additional (non-stellar) sources of heating, such as adiabatic
    compression, but is beyond the scope of this work.

The frequency with which the radiative transfer is calculated is
determined by the sound crossing time in the inner regions of the disc
as this is the time scale over which the scale height in this region
will vary and effect the transfer of radiation into the outer parts of
the disc.  The vertical sound crossing time at a radius of
10\,au is typically approximately 2 years, so a radiative transfer calculation is
performed after every four hydrodynamics time steps (equivalent to
every 2.4 years).

A cylindrical polar grid, with adaptive $r$ and $z$ cells was used for
the radiative transfer step.  The cells were split in the azimuthal
direction to give a cell spacing of 22.5 degrees. This azimuthal
splitting allows for the existence of non-axisymmetric instabilities
(such as radiatively driven warps) which may occur due to either
physical or numerical effects, while ensuring the calculation remains
tractable.  The AMR grid was constructed using a maximum mass per cell
of $3.0\times10^{25}\,\rm{g}$ and a density splitting condition of
$f_{\rm{split}}=0.1$. The density in a cylindrical region around the
source was reduced to a very low value to ensure that the disc did not
completely obscure the source.  The radius of this region is
$R_{\rm{gap}}=1.0\,\rm{au}$ which corresponds to the accretion radius
of the SPH sink particle. Two extra levels of grid refinement were
added in a box around the central source to ensure that the grid had
enough resolution to allow the gap around the source to be
resolved. The first box of extra refinement has size
$1\times10^{14}\,\rm{cm}$ with cells no larger than
$1\times10^{13}\,\rm{cm}$ and the second box has size
$4\times10^{13}\,\rm{cm}$ with cells no larger than
$4\times10^{12}\,\rm{cm}$.

Monte-Carlo radiative transfer calculations can be computationally
demanding so the calculation was parallelised using MPI (Message
Passing Interface) libraries. Parallelising code in this way allows
for parallel computation in both the radiation and hydrodynamics steps
and can be applied to a shared memory or a distributed memory computer
architecture. 
The Monte-Carlo calculation dominates the run time
    of the model but fortunately is very amenable to parallelisation, 
    due to the low communication overhead. The calculations were run
    on an SGI Altix ICE system using 16 compute nodes, each node
    comprising two quad core 2.83\,GHz Intel Xeon processors. A total
    of 128 MPI processes were used (one per core).
Evolving the disc using both the radiation and
hydrodynamics codes took a total wall time of 560 hours of which 440
hours was spent in radiation calculations.
The radiation calculations take approximately four times as long as the
hydrodynamics calculations so the combined code is significantly
slower than the SPH code alone. However the calculation is still
feasible and we expect the combined code only to be used in applications
where the increased accuracy of the Monte-Carlo radiation calculation,
over faster alternatives such as flux-limited diffusion, is of importance. 

\subsection{Results}

Plots of the disc internal energy are shown in Fig.~\ref{fig:log_u} at
six different times during the simulation. 
As these plots are in cylindrical polar co-ordinates, with
    different scales on the $r$ and $z$ axes, they have been shown as
    particle plots, rather than kernel integrated plots.
Figure~\ref{fig:frame1}
shows the initial state before the radiative transfer is switched on
when there is a vertically isothermal equation of state. The second
frame (Fig.~\ref{fig:frame2} is shortly after the radiative transfer
has been switched on. The inner disc has collapsed and there is
heating seen in parts of the disc exposed to radiation from the
central star. In Fig.~\ref{fig:frame3} a scale height enhancement is
present in the disc at a radius of approximately 20 au which shadows
the region immediately behind causing it to
cool. Figure~\ref{fig:frame4} shows several such scale height
enhancements and in each case there is cooling behind the enhancement
as the scale height fluctuations are able to affect the mid-plane
temperature due to vertical transport of radiation. 
In Fig.~\ref{fig:frame5} the inner region of the disc, within
    approximately 75\,au, has 
settled and scale height enhancements are seen only in the 
outer disc, implying an equilibrium time of order 1600~years at
75\,au. As the equilibrium time scales with radius as $r^\frac{3}{2}$ 
we expect the equilibrium time at 150\,au to be
of order 4500 years. By the last frame (Fig.~\ref{fig:frame6}) the
majority of the disc has reached a steady state in hydrostatic and
radiative equilibrium, indicating that it is not subject to
radiative-hydrodynamic instabilities.
\begin{figure*}
  \subfigure[]{\includegraphics[scale=0.3]{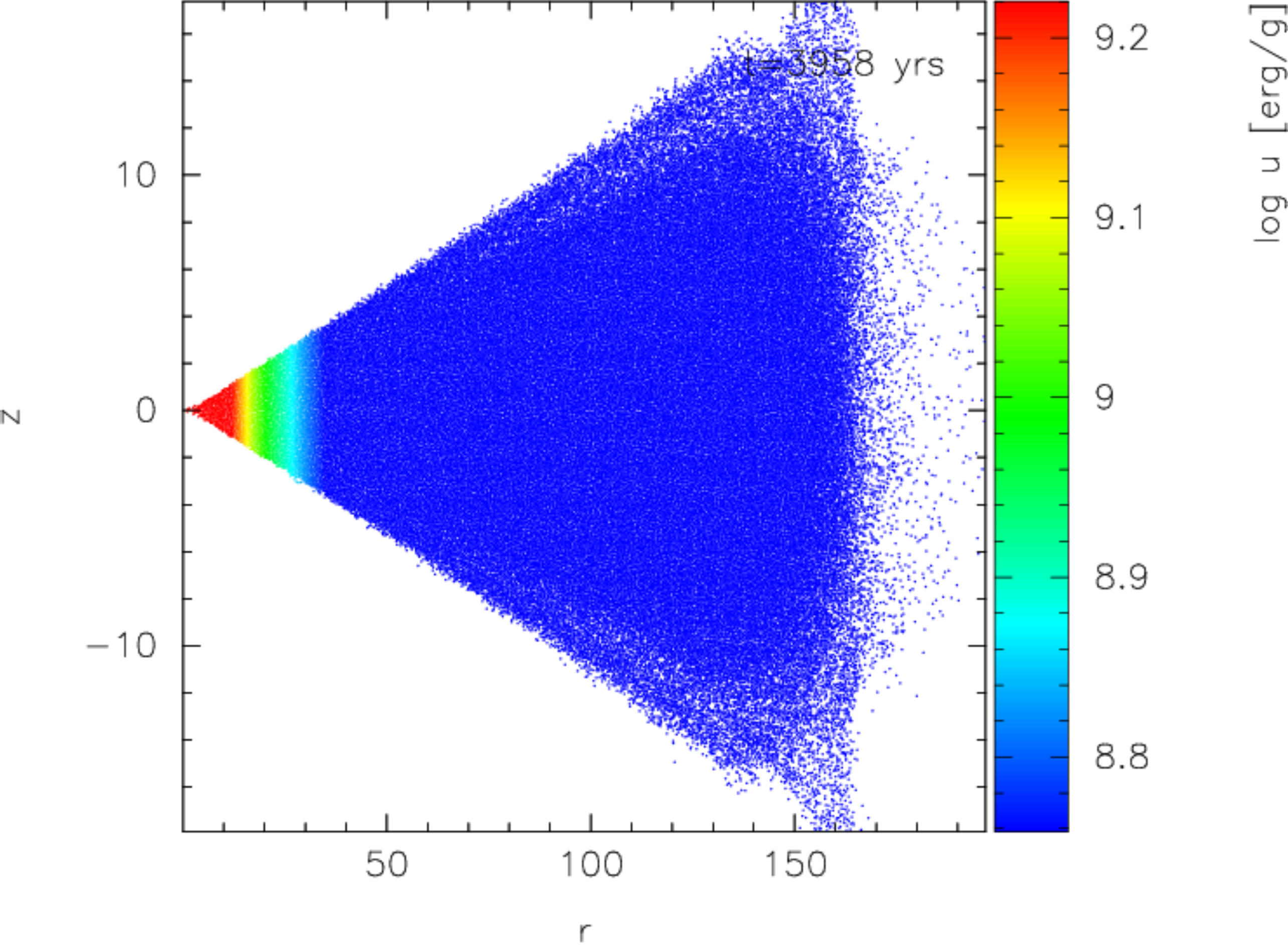}\label{fig:frame1}}
  \subfigure[]{\includegraphics[scale=0.3]{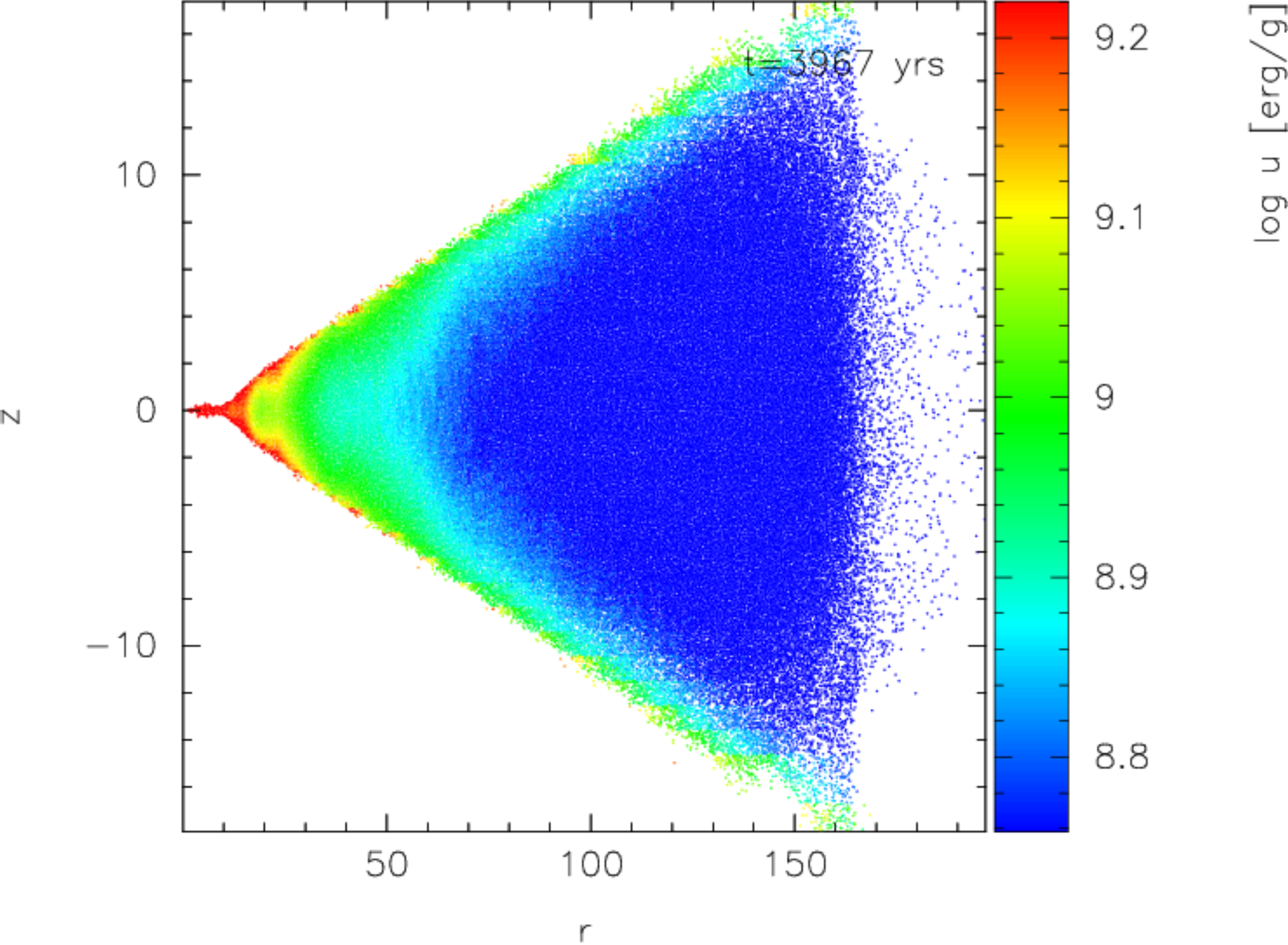}\label{fig:frame2}}
  \subfigure[]{\includegraphics[scale=0.3]{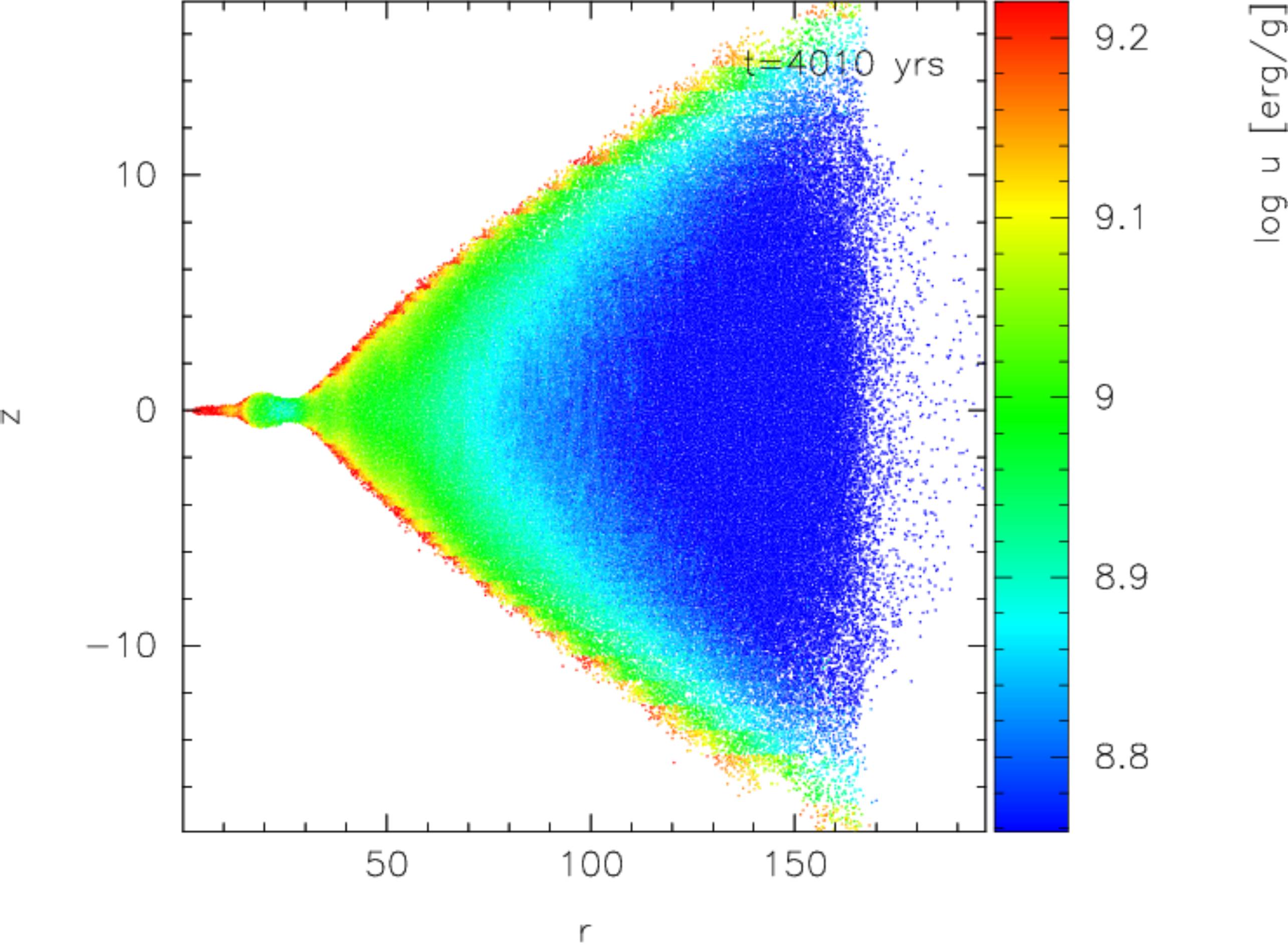}\label{fig:frame3}}
  \subfigure[]{\includegraphics[scale=0.3]{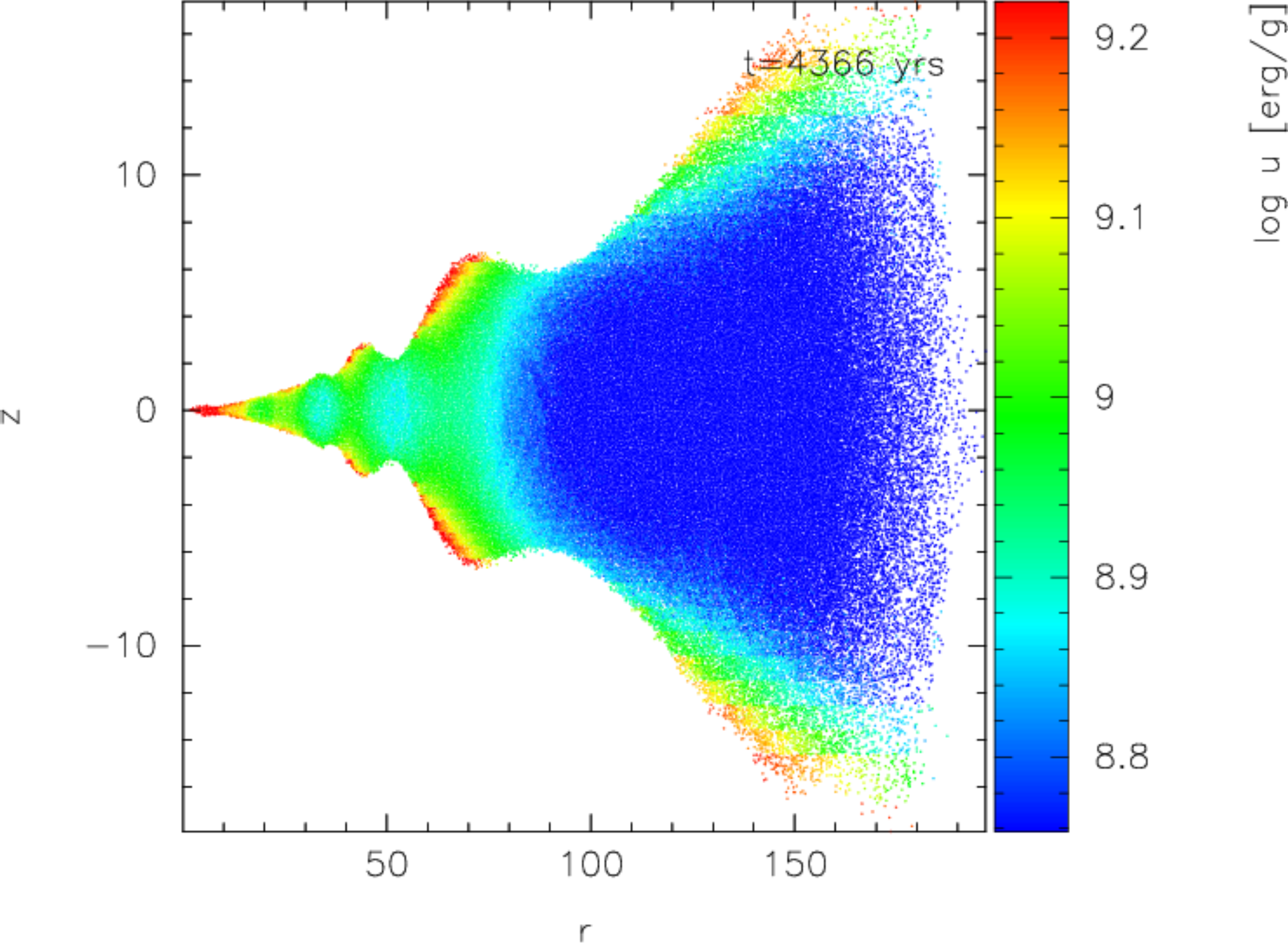}\label{fig:frame4}}
  \subfigure[]{\includegraphics[scale=0.3]{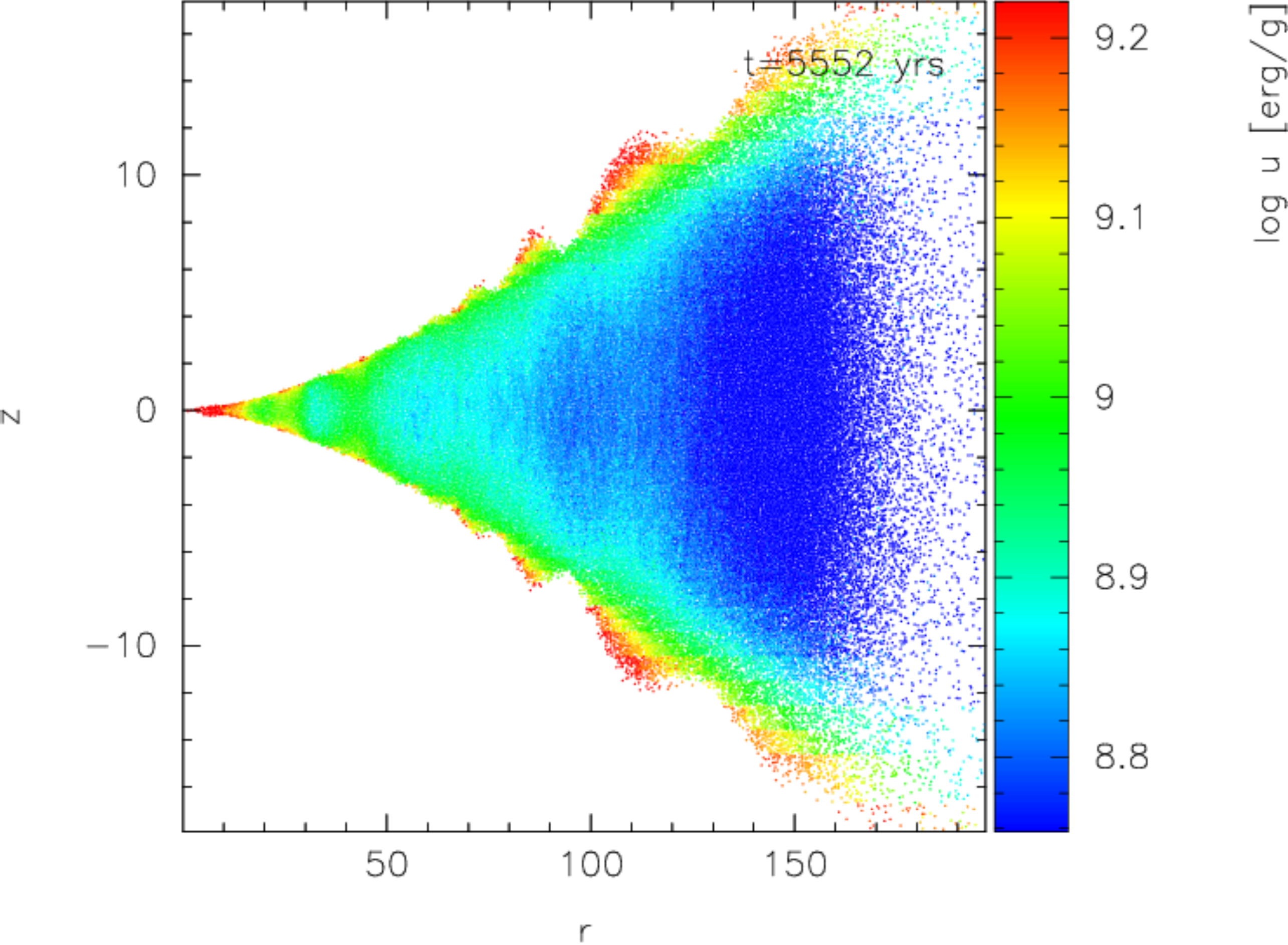}\label{fig:frame5}}
  \subfigure[]{\includegraphics[scale=0.3]{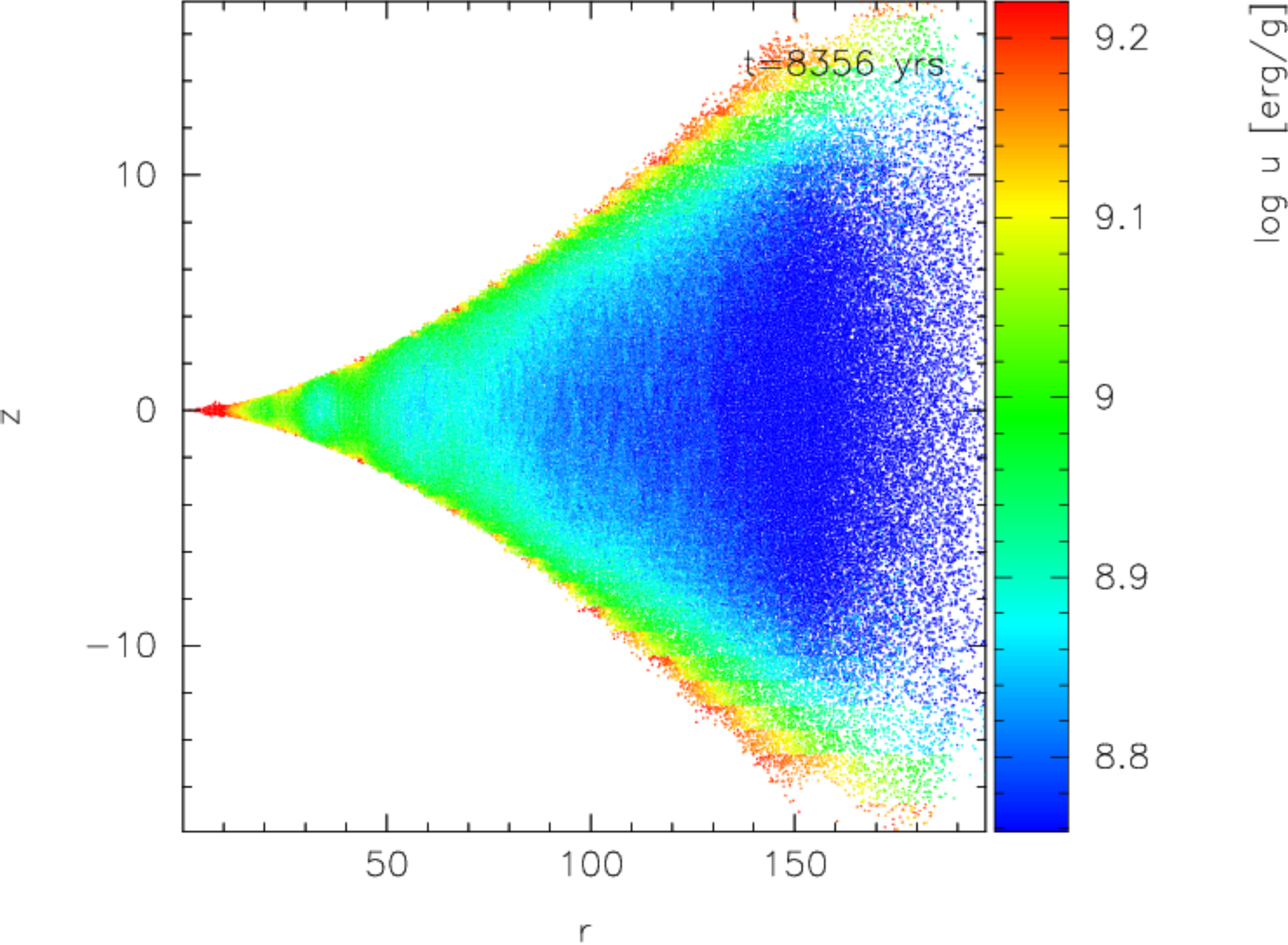}\label{fig:frame6}}
  \caption{Internal energy distribution in the model disc at six
    different times as the disc adjusts towards
    equilibrium. Figure~\ref{fig:frame1} shows the initial vertically
    isothermal state and Fig.~\ref{fig:frame6} shows the steady state in
    hydrostatic and radiative equilibrium. Intermediate frames show
    the effect of scale height fluctuations on the internal energy
    within the disc. Internal energy is plotted for each particle in a
    cylindrical polar co-ordinate system and the axis units are au.}
  \label{fig:log_u}
\end{figure*}

Transient oscillations, as the disc adjusts towards equilibrium, are
observed whether radiative transfer is included or not. They are
vertical motions which decay more rapidly at smaller radii and hence
give the appearance of propagating outwards, even though the motion is
vertical. Other disc models using radiation hydrodynamics have found
oscillatory solutions with inwards propagating waves
(e.g. \citealt{min_2009}) which are fundamentally different to the
transient effects seen in this case. However it should be noted that
\cite{min_2009} only found an oscillatory solution in their most
optically thick disc, which is significantly more massive than our
model disc, and may represent a different regime to that considered
here. We note that the disc retains its azimuthal symmetry throughout 
the simulation and we find no evidence for warping instabilities.

The surface density of the disc is shown in Fig.~\ref{fig:sigma}, the
scale height is shown in Fig.~\ref{fig:scale_height} and the radial
temperature profile is shown in Fig.~\ref{fig:mid_tem}.  The solid
lines are from the vertically isothermal disc shown in
Fig.~\ref{fig:frame1} and the dotted lines are the disc at the final
time step shown in Fig.~\ref{fig:frame6}. The scale height of the disc
was determined by allocating particles to 100 radial bins and fitting
a Gaussian to the vertical density profile; the scale height is
    taken to be the standard deviation of the fitted Gaussian. The radial temperature
profile is the average temperature within one scale height of the
mid-plane for each radial bin.
\begin{figure}
  \includegraphics[scale=0.3]{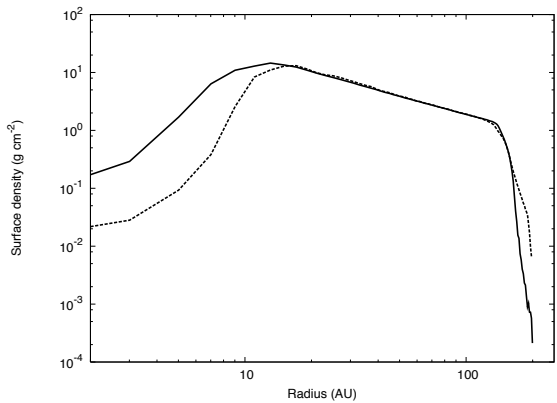}
  \caption{Surface density as a function of radius for the model
    disc. The solid line is the vertically isothermal disc prior to
    the radiative transfer code being switched on. The dotted line in
    shows the disc once is has reached a state of hydrostatic and
    radiative equilibrium.}
  \label{fig:sigma}
\end{figure}
\begin{figure}
  \includegraphics[scale=0.3]{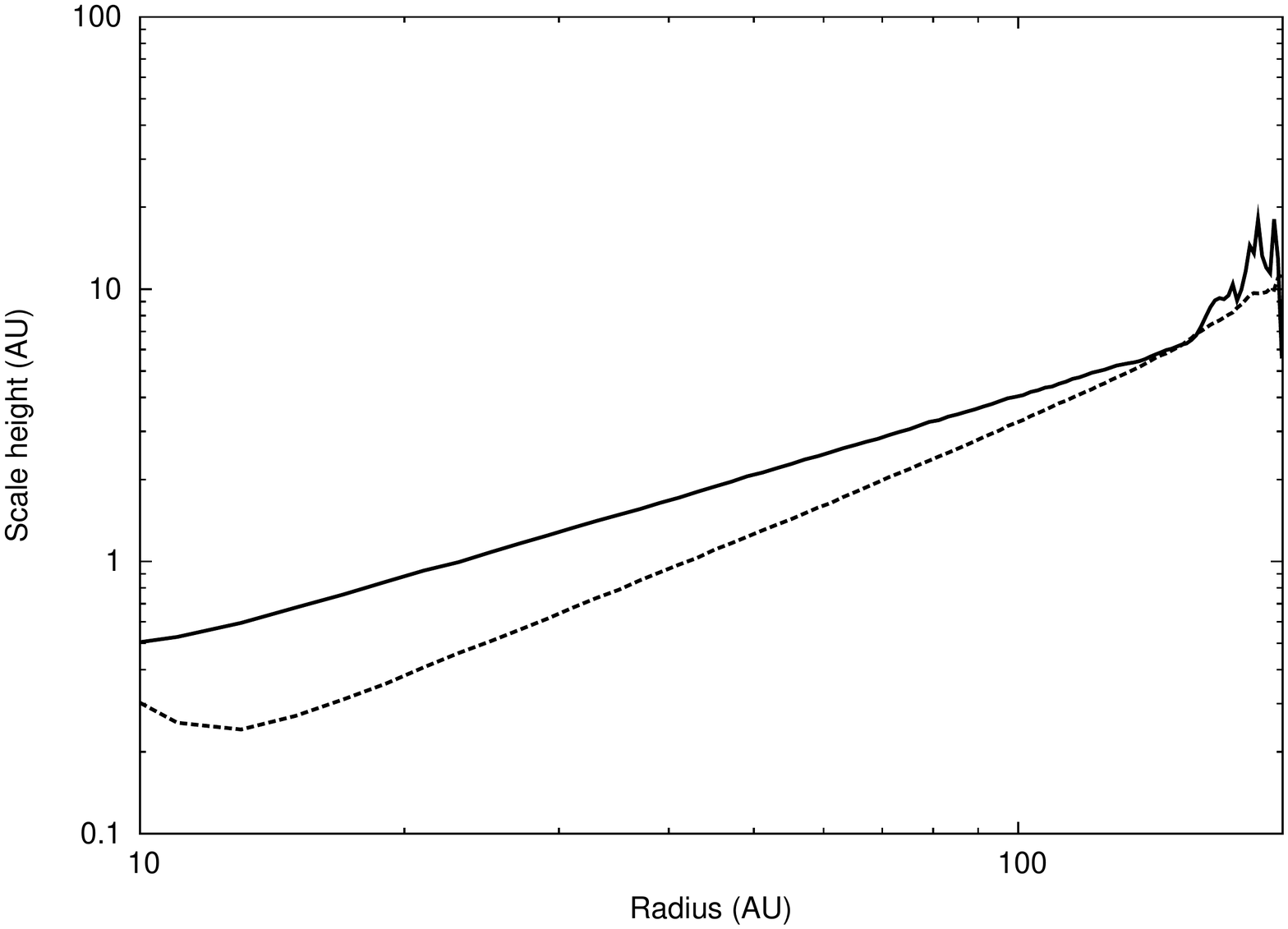}
  \caption{Scale height of the model disc as a function of radius. The
    solid line is the vertically isothermal disc prior to the
    radiative transfer code being switched on. The dotted line in
    shows the disc once is has reached a state of hydrostatic and
    radiative equilibrium.}
  \label{fig:scale_height}
\end{figure}
\begin{figure}
  \includegraphics[scale=0.3]{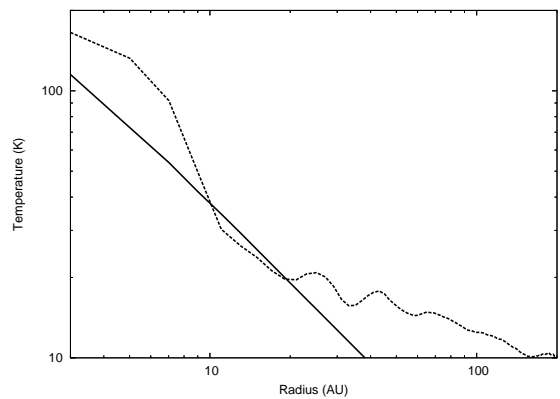}
  \caption{Average temperature within one scale height of the mid
    plane as a function of radius for the model disc. The solid line
    is the vertically isothermal disc prior to the radiative transfer
    code being switched on. The dotted line in shows the disc once is
    has reached a state of hydrostatic and radiative equilibrium.}
  \label{fig:mid_tem}
\end{figure}     
The majority of the surface density profile is well represented by a
power law and does not change significantly as a result of including
radiative transfer. Figures~\ref{fig:scale_height} and
\ref{fig:mid_tem} show that the disc is hotter close to the mid-plane,
except within a region between 10--20\,au of the centre, but has a
smaller scale height. Figure~\ref{fig:log_u} shows that the vertical
temperature gradient in the disc is such that temperature increases
with height out of the mid-plane. Although the disc is hotter when
radiative transfer is included the vertical temperature gradient tends
to reduce the pressure gradient which is required for hydrostatic
support resulting in a smaller scale height. 
A slight drop in scale height is evident, 
just outwards from the disc inner edge, which is indicative of a
puffed-up inner edge. \cite{dullemond_2001} find such a puffed-up
inner edge when the disc is truncated by dust
sublimation. Although the inner edge in our model is not due to dust
sublimation, and is found at a greater radius than the
\cite{dullemond_2001} case, we are still seeing a puffed up inner edge
due to irradiation of an optically thick inner rim.

{A notable feature in the surface density plot is the drop in
    surface density at small radii. This is due to the accretion of
    material by the central sink particle and will tend to move the
    optically thick inner edge out to larger
    radii. Figure~\ref{fig:central_density} shows the SPH density in
    the central region of the disc at the last time step of the
    simulation. The inner edge of the disc has moved outwards from its
initial value of 1\,au. For the purposes of this model we still have
an optically thick edge which will shadow the disc but the use of an
accreting sink particle is likely to be inappropriate if the exact
location of the disc inner edge needs to remain fixed.
\begin{figure}
  \includegraphics[scale=0.3]{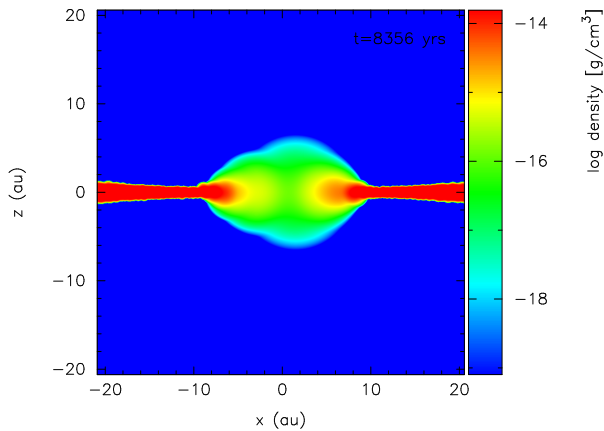}
  \caption{Kernel integrated plot of SPH density in the central region
    of the disc. This plot is from the last time step of the
    simulation and shows that the inner edge of the disc has moved
    outwards from its initial location at 1\,au due to the accretion
    of material by the sink particle. 
}
  \label{fig:central_density}
\end{figure}

\section{Conclusions}

\label{section:conclusions}

We have presented results from a hybrid SPH and AMR grid-based
radiative transfer code, which has been used to model a non-self
gravitating circumstellar disc. Validation tests of the method have
revealed a number of factors which must be taken into consideration
when setting up a model of this type.

Although we have found a method for constructing an AMR grid from SPH
particles which is robust, in terms of the total mass on the grid, we
found that it was necessary to add additional grid refinement in the
central region of the disc to allow the inner edge to be well
represented. The inner edge of the disc has a large
effect on the temperature distribution in the disc and on the emergent
SEDs. Consequently the accuracy of the solution is dependent upon the
accurate representation of a region which constitutes a small fraction
of the disc by volume. As the method combines both SPH and grid
elements it is important that both these components have sufficient
resolution to represent the disc inner edge. This presents a
particular challenge for a SPH code which uses equal mass particles
but is achievable with a large number of particles. Accurately
modelling the SED using this method is considerably more
demanding than accurately representing the temperature structure
within the disc. Temperature errors can be reduced to a manageable
level using a realistic number of SPH particles, yielding a 
temperature distribution suitable for use in a hydrodynamics
calculation, however even with a disc composed of $10^8$ SPH particles
the modelled SEDs show significant deviations from the benchmark results. 

The combined SPH-radiative transfer code was able to evolve a
circumstellar disc to a state of radiative and hydrostatic
equilibrium. As the disc settles towards an equilibrium state,
vertical motions are seen which give the appearance of an outward
propagating scale height fluctuation. This is a vertical motion which
decays more rapidly at smaller radii and is seen whether radiative
transfer is included or not. These are not the same as the
instabilities seen in other radiation hydrodynamics disc models
\citep{min_2009}. One consequence of these fluctuations is that they
can intercept radiation from the central star which can then be
vertically transported into the mid-plane of the disc hence even
thought the fluctuations are of a small amplitude they exert an
influence on the mid-plane temperature.

\section*{Acknowledgments}
Calculations presented here were performed using the University of
Exeter Supercomputer. DMA was funded by EPSRC grant EP/F011326/1. DAR
is funded by an STFC studentship. We thank Matthew Bate for useful
discussions. We would like to thank an anonymous referee for helpful
comments. 

\bibliographystyle{mn2e}
\bibliography{torus} 

\end{document}